\documentclass[traditabstract]{aa}
\usepackage{graphicx}
\usepackage{txfonts}
\newcommand{\degree}{^\circ}
\usepackage{float}

\begin{document}

   \title{V444 Cyg X-ray and polarimetric variability:\\ Radiative and Coriolis forces shape the wind collision region}


   \author{J. R. Lomax\inst{1,2,3}
          \and
          Y. Naz\'{e}\inst{1}\fnmsep\thanks{Research Associate FRS-FNRS}
          \and
          J. L. Hoffman\inst{2}
          \and
          C. M. P. Russell\inst{4}\fnmsep\thanks{Current affiliation: Department of Physics \& Astronomy, University of Delaware}
          \and
          M. De Becker\inst{1}
          \and
          M. F. Corcoran \inst{5}
          \and
          J. W. Davidson\inst{6}
          \and
          H. R. Neilson\inst{7}
          \and
          S. Owocki\inst{8}
          \and
          J. M. Pittard\inst{9}
          \and
          A. M. T. Pollock\inst{10}
          }

   \institute{ D\'{e}partement AGO, Universit\'{e} de Li\`{e}ge, All\'{e}e du 6 Ao\^{u}t 				17, B\^{a}t. B5C, B4000-Li\`{e}ge, Belgium\\
              \email{Jamie.R.Lomax@ou.edu}
              \email{naze@astro.ulg.ac.be}
         \and
          Department of Physics \& Astronomy, University of Denver, 2112 E. Wesley Ave., Denver, CO, 80210, USA
         \and
               Homer L. Dodge Department of Physics \& Astronomy, University of Oklahoma, 440 W Brooks Street, Norman, OK 73019, USA
             \and
        Faculty of Engineering, Hokkai-Gakuen University, Toyohira-ku, Sapporo 062-8605, Japan
         \and 
         	CRESST/NASA Goddard Space Flight Center, Code 662, Greenbelt, MD, 20771; University Space Research Association, Columbia, MD, USA
         \and
         Department of Physics \& Astronomy, University of Toledo, 2801 W Bancroft St, Toledo, OH, 43606, USA
         \and
         	Department of Physics \& Astronomy, East Tennessee State University, Box 70652, Johnson City, TN, 37614, USA
        \and
        Department of Physics \& Astronomy, University of Delaware, Newark, DE, 19716, USA
        \and
        School of Physics \& Astronomy, The University of Leeds, Leeds, LS2 9JT, UK 
        \and
        European Space Agency, \textit{XMM-Newton} SOC, ESAC, Apartado 78, 28691, Villanueva de la Ca\~{n}ada, Madrid, Spain
             }

   \date{Received }

 
  \abstract
  {We present results from a study of the eclipsing, colliding-wind binary V444 Cyg that uses a combination of X-ray and optical spectropolarimetric methods to describe the 3-D nature of the shock and wind structure within the system. We have created the most complete X-ray light curve of V444 Cyg to date using 40 ksec of new data from \textit{Swift}, and 200 ksec of new and archived \textit{XMM-Newton} observations. In addition, we have characterized the intrinsic, polarimetric phase-dependent behavior of the strongest optical emission lines using data obtained with the University of Wisconsin's Half-Wave Spectropolarimeter. We have detected evidence of the Coriolis distortion of the wind-wind collision in the X-ray regime, which manifests itself through asymmetric behavior around the eclipses in the system's X-ray light curves. The large opening angle of the X-ray emitting region, as well as its location (i.e. the WN wind does not collide with the O star, but rather its wind) are evidence of radiative braking/inhibition occurring within the system. Additionally, the polarimetric results show evidence of the cavity the wind-wind collision region carves out of the Wolf-Rayet star's wind.}


   \maketitle
%

\section{Introduction}
V444 Cygni (also known as WR 139 and HD 193576) is one of the few known eclipsing Wolf-Rayet (WR) binary systems with colliding winds and a circular orbit \citep[\textit{i} = 78.3$\degree \pm$ 0.4$\degree$]{Eris}. Its distance has been disputed. The Hipparcos measured parallax gives a distance of 0.6 $\pm$ 0.3 kpc \citep{Leeuwen}, although \cite{Kron}, \cite{Forbes}, and \cite{Nugis} find distances between 1.15 kpc and 1.72 kpc. At any of these distances, it is the closest known example of an eclipsing WR binary. By convention the star that was initially more massive in a WR+O system is defined as the primary. However, in order to agree with previous optical and infrared (IR) studies which defined the primary as the brighter of the two stars, we here denote the main-sequence O6 star as the primary and the WN5 star as the secondary. Table \ref{SysParam} lists the V444 Cyg system parameters. 

Evidence of colliding winds within the V444 Cyg system is considerable and comes from several wavelength regimes. Studies using \textit{IUE} spectra used variations in the terminal velocity and material density inferred from complex emission features to diagnose the presence of colliding winds \citep{Koen,Shore}. Additional evidence from \textit{Einstein}, \textit{ROSAT}, \textit{ASCA}, and \textit{XMM-Newton} suggests that at least part of the observed X-ray emission is due to the wind collision region; the measured X-ray temperature is higher than expected for single WR and O stars, and the variability is consistent with a colliding wind scenario \citep{Moffat, Pollock, Corcoran, Maeda, Bhatt, Fauchez}. 

Today V444 Cyg is considered the canonical close, short-period, colliding-wind binary system. It is the example system for both radiative inhibition and radiative braking because of the system's small orbital separation $a=35.97$ R$_\sun$ \citep{Stevens1994,Owocki,Eris}. Radiative inhibition is a process by which the acceleration of a wind is reduced by the radiation from a companion star. By contrast, radiative braking describes a scenario in which a wind is slowed after reaching large velocities by the radiation from a companion. However, despite the system's brightness, the phase coverage of X-ray observations of V444 Cyg has been insufficient to place reliable constraints on these processes until now.

\cite{Bhatt} and \cite{Fauchez} have both analyzed the 2004 \textit{XMM-Newton} observations of V444 Cyg, which covered only half of the system's cycle. Analysis by \cite{Fauchez} suggested some unexpected results concerning the wind collision region. These authors found that the hard X-ray emitting region may be positioned close to the WN star, suggesting that strong RIB processes may occur within the system \citep{Owocki,Stevens1994}. We obtained new \textit{XMM-Newton} observations to complete the X-ray light curve and confirm this interpretation. In addition, Coriolis distortion due to the orbital motion of the stars may affect the shape and orientation of the wind-wind interaction region. Our new observations test this prediction by investigating whether the two halves of the light curve mirror each other. Any asymmetries in the light curve can be used to construct a quantitative model of the wind collision region and provide important new information about its structure. 

\begin{center}
   \begin{table}
      \caption[]{V444 Cyg System Parameters}
         \label{SysParam}
   
        \begin{tabular}{llll}
            \hline
            \hline
            Parameter & WN star & O star & References\\
            \hline

    R (R$_\sun$) & 2.9 & 6.85 & 1,2\\
    M (M$_\sun$) & 12.4 & 28.4 & 3\\
    $\dot{M}$ (M$_\sun$ yr$^{ -1}$) & $6.76\times 10^{ -6}$ & $5.8\times 10^{ -7}$ & 3\\
    $v_{\infty}$ (km s$^{-1}$) & 2500 & 1700 & 4\\
            \hline
         \end{tabular}\\
         \tablefoot{References: 1 \citep{Corcoran}; 2 \citep{Eris}; 3 \citep{Hirv}; 4 \citep{Stevens}}
   \end{table}
\end{center} 

Optical spectropolarimetric observations can place additional constraints on the geometry of circumstellar material within the system. Light scattering from free electrons in the ionized circumstellar material is responsible for the phase-dependent polarization observed in V444 Cyg \citep{Robert,StLouis}. Since electron scattering preserves geometric information about the scattering region, analyzing the polarization behavior of the system as a function of wavelength and orbital phase allows us to describe the scattering regions that produce the polarization in different spectral features. In the \textit{UBVRI} bands, the observed phase-locked linear polarization variations are dominated by the O star's occultation of photons originating from the WR star and scattering in a region of varying electron density \citep{StLouis}. \cite{StLouis} also found that the polarization behavior of the system near secondary eclipse deviates from the theoretical predictions of \cite{BME}, possibly due to the WR wind's distortion from spherical symmetry as a result of the binary's orbital motion. \cite{Kurosawa} modeled the continuum polarization and found that they could reproduce the observations with the WR wind alone; the presence of the O-star wind and the wind-wind collision region do not affect the continuum polarization.

In this paper, we present the results of four new \textit{XMM-Newton} observations, which we combine with six archival \textit{XMM-Newton} observations to construct an X-ray light curve of the system with the best coverage to date. We supplement the \textit{XMM-Newton} data with new observations from \textit{Swift} that cover both eclipses. The optical and infrared primary (WN star in front of the O star) and secondary eclipses (O star in front of the WN star) are well covered with our combined data sets. We also present new polarization curves of the V444 Cyg system in several strong optical emission lines, using data obtained with the University of Wisconsin's Half-Wave Spectropolarimeter (HPOL) at the Pine Bluff and Ritter Observatories. Our resulting multi-wavelength, multi-technique study provides new information about the structure of the winds (through spectropolarimetry) and wind-wind interaction region (through X-rays), which we use to infer details about the 3D nature of the shock and wind structure. In Section 2 we describe our observations. Section 3 presents our observational results, including X-ray light curves, X-ray spectra, and line polarization variations. In this section we also discuss initial interpretations. We analyze our findings in Section 4 using simple modeling techniques and summarize our conclusions in Section 5.


\section{Observations}

This study uses data from several distinct data sets. Our first set of data consists of ten X-ray observations taken during two different years with the \textit{XMM-Newton} European Photon Imaging Cameras (EPIC); the second set is 40 ksec of observations of V444 Cyg taken over two weeks by \textit{Swift}. The third data set consists of 14 observations taken with the University of Wisconsin's Half-Wave Spectropolarimeter (HPOL) at the 0.9m telescope at Pine Bluff Observatory (PBO); and the last includes 6 observations obtained with HPOL at the 1.0m telescope at the University of Toledo's Ritter Observatory. We phased all observations using the ephemeris given by \cite{Eris} $$T = \textrm{HJD\space} 2441164.311 + 4.212454 E,$$ where $E$ is the decimal number of orbits of the system since the primary eclipse, when the O star is eclipsed by the WN star, that occurred on HJD $2,441,164.311$. 

\subsection{\textit{XMM-Newton}}

Our ten \textit{XMM-Newton} observations were taken using the EPIC instrument in full frame mode with the medium filter. The first six of those observations are from 2004 while the last four were taken in 2012. The lengths of the observations vary; see Table \ref{XMMInfo} for their observation ID numbers, revolution numbers, start times, durations, and phase ranges. In total they cover approximately 65\% of the orbit.

We reduced all of these observations using version 12.0.1 of the \textit{XMM-Newton} SAS software. Because of the faintness of the source, pile up is not a problem; however, background flares were excised from several observations (see Table \ref{XMMInfo}). Additionally, data from both MOS CCDs for observation 0206240401 (revolution number 0819) were not usable due to a strong flare affecting both MOS1 and MOS2 exposures. For each \textit{XMM-Newton} observation, we extracted spectra for all EPIC cameras in a circular region with a 30" radius around the source such that oversampling is limited to a factor of five and the minimum signal-to-noise ratio per bin is three. We extracted the background from a nearby area that was 35" in radius and devoid of X-ray sources. Additionally, we extracted light curves for these regions, correcting the data for parts of the point spread function not in the extraction region (using SAS task {\tt{epiclccorr}}) and correcting to equivalent on-axis count rates. We created light curves for each observation in the soft (0.4-2.0 keV), medium (2.0-5.0 keV), hard (5.0-8.0 keV), and total (0.4-10.0 keV) bands.

   \begin{table*}[ht]  
   \begin{center}
      \caption[]{\textit{XMM-Newton} Observation Information for V444 Cyg}
         \label{XMMInfo}
   
        \begin{tabular}{llllll}
            \hline
            \hline
            Obs. ID      &  Rev. & Start Time (HJD) & Duration (s) & Phase Range \tablefootmark{a} & Flare?\\
            \hline
            0206240201 & 0814   &  2453144.997     & 11672 & 0.112-0.171& No \\
    0206240301 & 0818   &  2453152.986     & 11672 & 0.011-0.070& No \\
    0206240401\tablefootmark{b} & 0819   &   2453154.986    & 10034 & 0.485-0.545& Yes \\
    0206240501 & 0823   &  2453162.969     & 11672 & 0.385-0.444& No \\
    0206240701 & 0827   &  2453170.943     & 11672 & 0.284-0.343& Yes \\
    0206240801 & 0895   &  2453306.476     & 19667 & 0.450-0.510& Yes \\
    0692810401 & 2272  &   2456053.363    & 45272 & 0.505-0.683& No \\
    0692810601 & 2275  &   2456059.025    & 18672 & 0.870-0.929& Yes \\
    0692810501 & 2283  &   2456075.221    & 13672 & 0.727-0.787& Yes \\
    0692810301 & 2292  &   2456093.118    & 50067 & 0.941-0.119& No \\
    
            \hline
         \end{tabular}\\
         \tablefoot{$^{(a)}$ Phases were calculated using the ephemeris from \cite{Eris}. $^{(b)}$Data from MOS1 and MOS2 were lost for this observation due to a strong flare.}
     \end{center}
   \end{table*} 

\subsection{\textit{Swift}}

\textit{Swift} observed V444 Cyg for 40 ks over the course of two weeks in 2011 with the XRT instrument (see Table \ref{SwiftInfo}). Because observations were not continuous, we list only the start time and the total exposure for each observation ID. We processed the \textit{Swift} data with the online tool at the UK \textit{Swift} Science Data Center\footnote{http://www.swift.ac.uk/user\_objects/}, which produced a concatenated light curve with 100s bins, built for the soft (0.4-2.0 keV) and total (0.4-10.0 keV) bands. The small bins were then aggregated into 8 phase bins (0.00-0.5, 0.05-0.10, 0.40-0.45, 0.45-0.50, 0.50-0.55, 0.55-0.60, 0.90-0.95, and 0.95-1.00). Our observations have no pile up because count rates are low (between 0.01 and 0.04 cts s$^{ -1}$).

\begin{center}
   \begin{table}[t]
      \caption[]{\textit{Swift}-XRT Observation Information for V444 Cyg}
         \label{SwiftInfo}
   
        \begin{tabular}{lll}
            \hline
            \hline
            Observation ID & Start Time (HJD) & Total Exposure (s)\\
            \hline

    31983002 & 2455728.619 & 8970\\
    31983003 & 2455730.627 & 8969\\
    31983004 & 2455732.684 & 8048\\
    31983005 & 2455734.778 & 6749\\
    31983006 & 2455735.501 & 3277\\
    31983007 & 2455736.978 & 4677\\
    31983008 & 2455739.305 & 2646\\
    31983009 & 2455743.586 & 2270\\
            \hline
         \end{tabular}\\
   \end{table}
\end{center} 

\vspace{-40pt}
\subsection{HPOL}
Our HPOL data can be divided into two subgroups. Taken between 1989 October and 1994 December, the first 14 HPOL observations used a Reticon dual photodiode array detector with a wavelength range of 3200-7600 \AA\space and a resolution of 25 \AA\space (see Wolff et al. 1996 for further instrument information). During this period HPOL was at the Pine Bluff Observatory (hereafter HPOL@PBO). The last six observations were conducted between 2012 May and 2012 December with the refurbished HPOL at Ritter Observatory (hereafter HPOL@Ritter; see Davidson et al. 2014 for HPOL@Ritter instrument information). These four observations used a CCD-based system with a wavelength range of 3200 \AA -10500 \AA , and a spectral resolution of 7.5 \AA\space below 6000 \AA\space and 10 \AA\space above \citep{NordsieckHarris}. 

Table \ref{HPOLInfo} lists the orbital phase along with the civil and heliocentric Julian dates for the midpoint of each HPOL observation. All of the HPOL@PBO observations covered the full spectral range of the Reticon detector system. Only one of the HPOL@Ritter observations (2012 Oct 22) covered the full spectral range of the CCD detector system; the others used only the blue grating (3200 \AA -6000 \AA). Each HPOL observation typically lasted between one and three hours (approximately 0.010 to 0.030 in phase). We reduced the HPOL data using the REDUCE software package which is specific to HPOL and described by Wolff et al. (1996). Astronomers at PBO and Ritter Observatory determine instrumental polarization for HPOL by periodically analyzing observations of unpolarized standard stars \citep{Davidson}. We removed this contribution from the data as part of the reduction process.

\begin{center}
   \begin{table*}[t]
   \begin{center}
      \caption[]{HPOL Observation Information for V444 Cyg}
         \label{HPOLInfo}
   
        \begin{tabular}{lllll}
            \hline
            \hline
            Date & Observatory & Detector & Midpoint HJD & Phase\tablefootmark{a}\\
            \hline

    1989 Oct 03 & PBO   & Reticon & 2447802.15 & 0.790 \\
    1990 Jul 24 & PBO   & Reticon & 2448096.26 & 0.583 \\
    1990 Oct 23 & PBO   & Reticon & 2448187.15 & 0.185 \\
    1990 Oct 24 & PBO   & Reticon & 2448188.10 & 0.363 \\
    1991 Aug 21 & PBO   & Reticon & 2448489.20 & 0.878 \\
    1991 Aug 31 & PBO   & Reticon & 2448499.31 & 0.252 \\
    1991 Oct 10 & PBO   & Reticon & 2448539.21 & 0.747 \\
    1991 Nov 17 & PBO   & Reticon & 2448577.11 & 0.708 \\
    1994 Jun 15 & PBO   & Reticon & 2449518.36 & 0.153 \\
    1994 Jul 03 & PBO   & Reticon & 2449536.35 & 0.426 \\
    1994 Aug 16 & PBO   & Reticon & 2449580.19 & 0.872 \\
    1994 Oct 06 & PBO   & Reticon & 2449631.23 & 0.978 \\
    1994 Nov 12 & PBO   & Reticon & 2449668.12 & 0.702 \\
    1994 Dec 01 & PBO   & Reticon & 2449687.12 & 0.212 \\
    2012 May 11\tablefootmark{b} & Ritter & CCD   & 2456058.87 & 0.810 \\
    2012 Jul 13\tablefootmark{b} & Ritter & CCD   & 2456121.83 & 0.767 \\
    2012 Jul 14\tablefootmark{b} & Ritter & CCD   & 2456122.87 & 0.004 \\
    2012 Oct 22 & Ritter & CCD   & 2456222.58 & 0.684 \\
    2012 Oct 26\tablefootmark{b} & Ritter & CCD   & 2456226.69 & 0.692 \\
    2012 Dec 14\tablefootmark{b} & Ritter & CCD   & 2456275.55 & 0.265 \\
            \hline
         \end{tabular}\\
	\tablefoot{$^{(a)}$ Phases were calculated using the ephemeris in \cite{Eris}. $^{(b)}$ These observations used only the blue grating (3200 -- 6000 \AA). All other HPOL observations are full spectrum for their respective detectors.}         

	\end{center}
   \end{table*} 
\end{center}

\section{Results}

\subsection{\textit{XMM-Newton} and \textit{Swift} light curves}
\begin{center}
\begin{figure*}[p]
\includegraphics[scale=1.0]{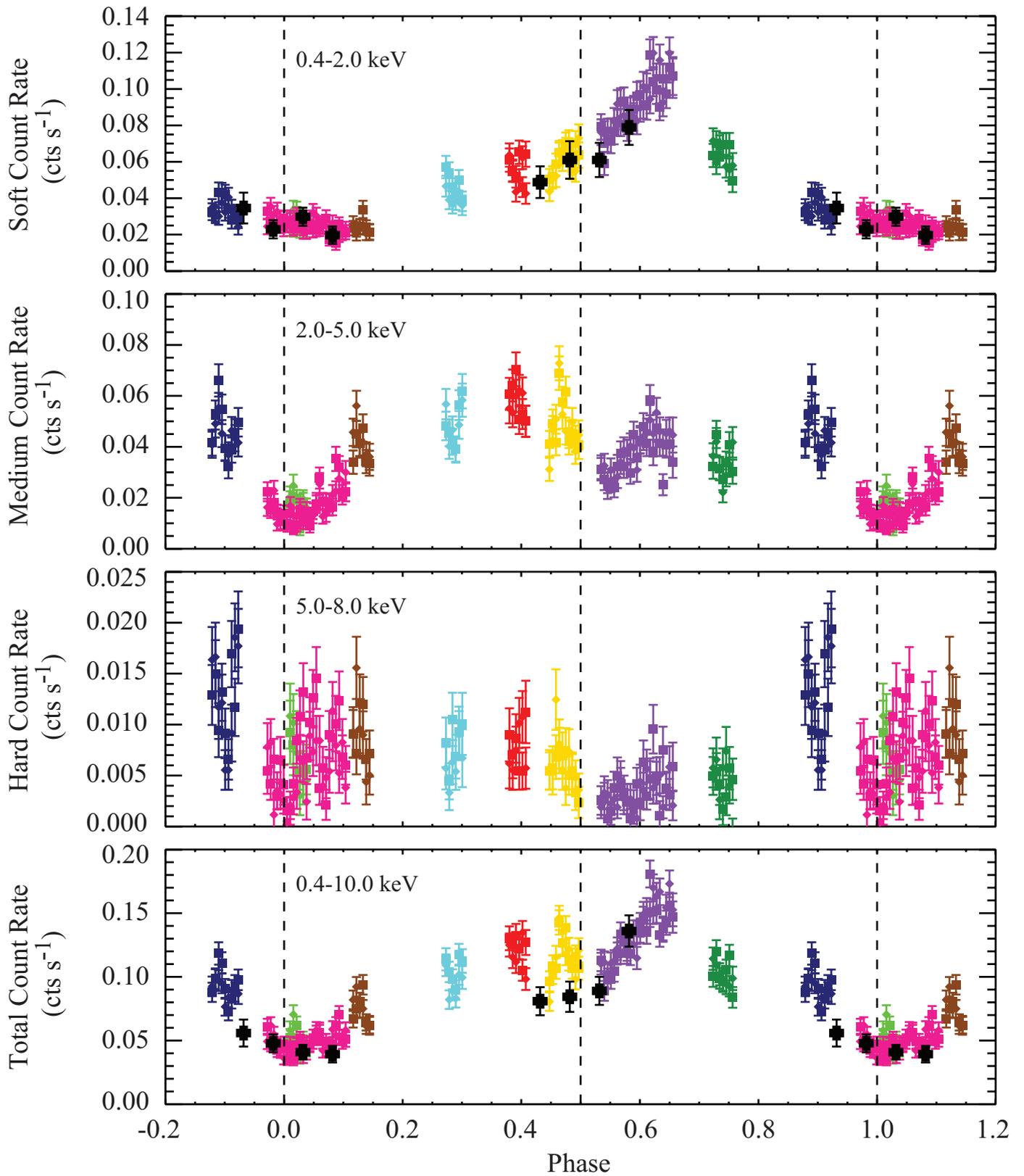}
\caption{X-ray count rates from the \textit{Swift} (black crosses), \textit{XMM-Newton} MOS1 (diamonds), and MOS2 (squares) observations discussed in Section 3.2. Colors indicate data from different \textit{XMM-Newton} observations: revolution number 0814=brown; 0818=green; 0819=blue; 0823=red; 0827=light blue; 0895=yellow; 2272=purple; 2275=dark blue; 2283=dark green; and 2292=pink. \textit{Swift} data have been converted into an \textit{XMM-Newton} equivalent count rate using the WebPIMMS software package. \textit{From top:} Count rate in the soft (0.4-2.0 keV), medium (2.0-5.0 keV), hard (5.0-8.0 keV), and total (0.4-10.0 keV) bands versus phase. All data have been wrapped in phase so that more than one complete cycle is shown. The dotted vertical lines represent phases 0.0, 0.5, and 1.0.}\label{XrayLCMOS}
\end{figure*}
\end{center}

\begin{center}
\begin{figure*}[!ht]
\includegraphics[scale=1.0]{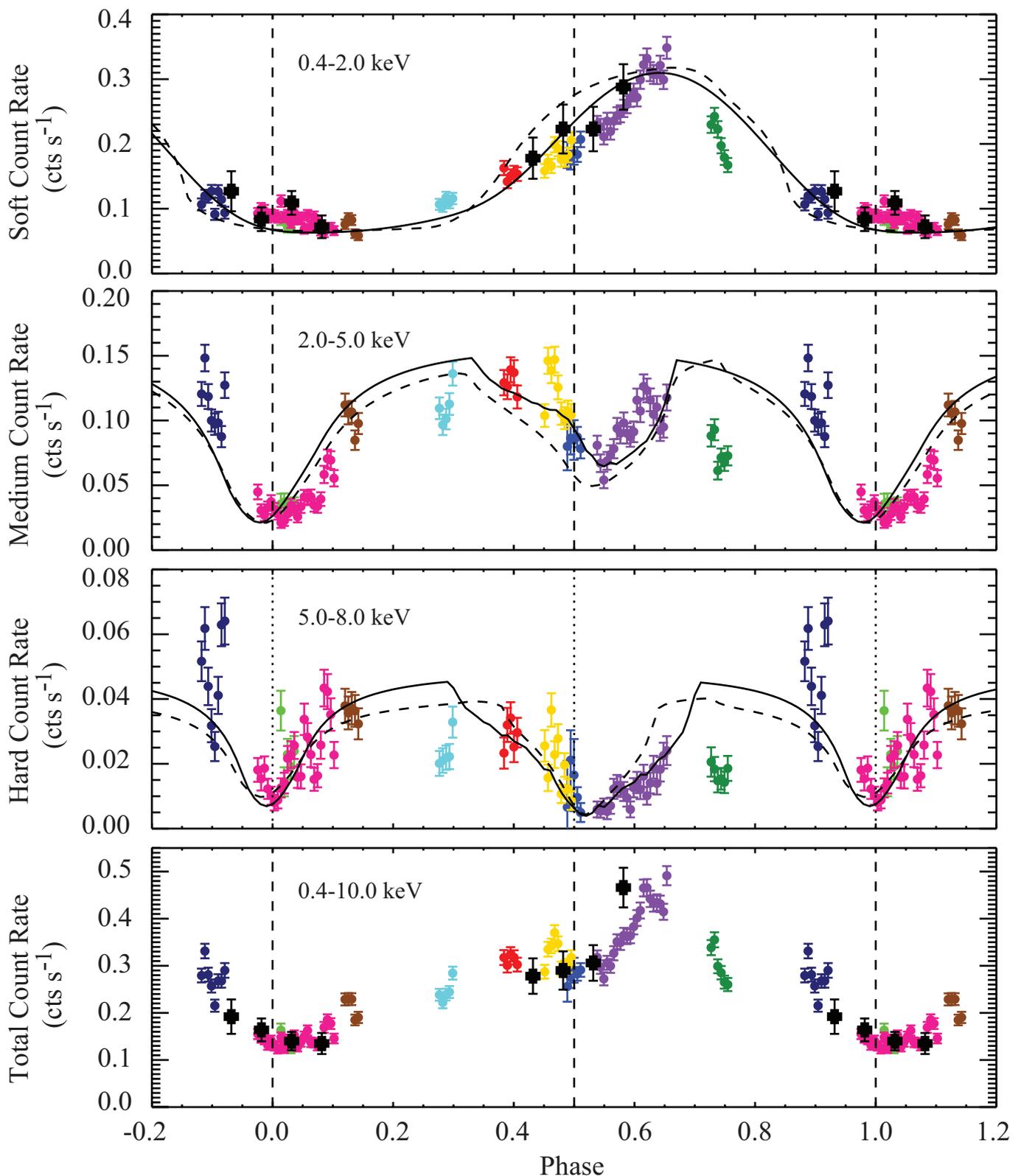}
\caption{Same as Figure \ref{XrayLCMOS}, but for the \textit{XMM-Newton} PN camera. The solid lines in the top three panels represent the results of our occultation plus WN-wind absorption model fits, while the dashed lines represent simulated light curves for the two-wind plus shock cone models (Section 4.1).}\label{XrayLCPN}
\end{figure*}
\end{center}

Our extracted light curves are displayed in Figures \ref{XrayLCMOS} (MOS1 and 2) and \ref{XrayLCPN} (PN) with 2 ks binning in the following bands: soft (0.4-2.0 keV), medium (2.0-5.0 keV), hard (5.0-8.0 keV), and total (0.4-10.0 keV). We removed bins with a fractional exposure less than 0.5 due to low signal-to-noise ratios. In addition, the MOS data from observation 0206240401 (revolution 0819) are not shown in any band because they were lost due to a strong flaring event. The PN data have a higher count rate than the MOS data due to the higher sensitivity of the PN camera. We converted the count rates of our \textit{Swift} observations into \textit{XMM-Newton} equivalent rates using the WebPIMMS software package and we overplot them in the soft and total bands (Figures \ref{XrayLCMOS} and \ref{XrayLCPN}). Although we have \textit{Swift} data in the medium and hard bands, we do not plot those data here due to their large uncertainties. These represent the most complete X-ray light curves of the V444 Cyg system to date. Additionally, they are the first to cover both the ingress and egress of the system's X-ray eclipses (phase 0.0 and 0.55), which allowed us to place important constraints on the size and location of the X-ray emitting region (Section 4). We describe the behaviors of each individual band below.

In the soft-band light curve, the minimum count rate occurs between phases 0.1 and 0.2, while the maximum occurs near phase 0.63 (top panels of Figures \ref{XrayLCMOS} and \ref{XrayLCPN}). The increase from minimum to maximum and the decrease from maximum to minimum are smooth; however, the count rate increase to the maximum occurs much more slowly than the subsequent decrease. This may indicate that the leading shock edge is brighter than the trailing edge. However, our data do no completely cover the decline.

The medium-band and hard-band light curves (second and third panels of Figures \ref{XrayLCMOS} and \ref{XrayLCPN}) show eclipses with minima near phase 0.0 (primary eclipse) and 0.55 (secondary eclipse). Although the primary eclipse is symmetric in phase, the secondary eclipse is not; it starts at approximately phase 0.47 but ends at phase 0.63. Therefore, the system enters secondary eclipse more quickly than it recovers from it. Additionally, the count rate is higher just before secondary eclipse than just afterward.

Since the medium and hard X-rays likely come from higher temperature gas than soft X-rays and the eclipses of the gas occur near the same phases as the optical eclipses, this behavior can be explained by physical occultation effects. That is, the eclipses in the medium and hard X-rays occur when the stars occult hot plasma in the wind-wind collision in and around the stagnation point, where the two winds collide head on. However, the secondary eclipse is broader in phase than the O star (approximately 0.1) so either the hard X-ray emitting region is large and never fully eclipsed, which is consistent with the count rate never dropping to zero, or the O-star wind is also responsible for a portion of the eclipse. In the latter less likely case, the count rate may never drop to zero because photons at these energies are not as readily absorbed within winds (see Marchenko et al. 1997 and Kurosawa et al. 2001). The asymmetry in the eclipse near phase 0.55 can  be explained if the hard X-ray emitting region does not lie on the line of centers connecting the two stars due to Coriolis distortion of the wind-wind interaction region (see Section 4.1 for the implications of this scenario). Additionally the broad and deep (but nonzero) primary eclipse suggests that the WN star appears larger than what is suggested by visual and ultraviolet light curve analysis \citep{Chere,StLouis}. The dense wind of the WN star gives rise to a wind eclipse that is broader than the eclipse created by the stellar surface alone.

\begin{center}
\begin{figure*}[]
\includegraphics[scale=0.5]{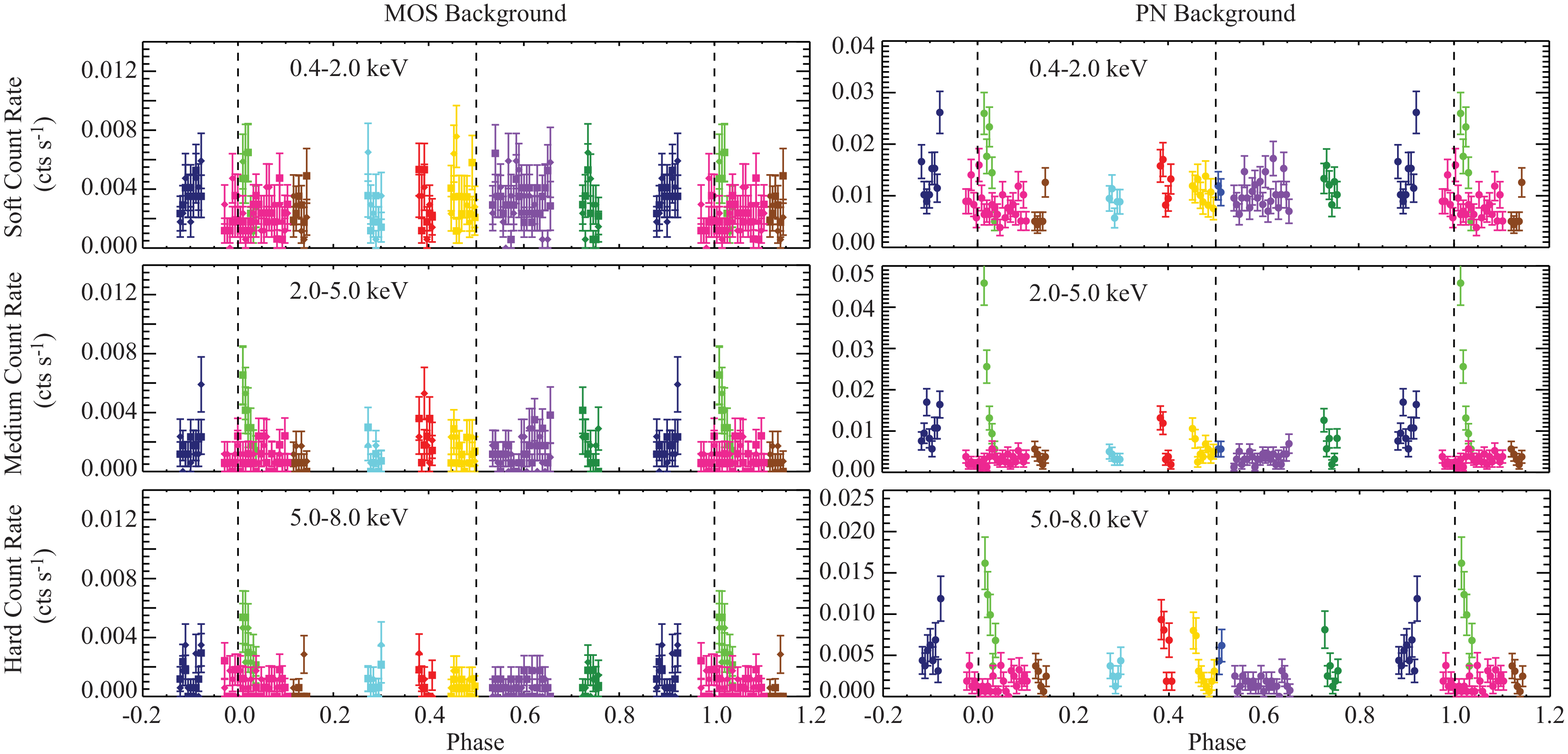}
\caption{X-ray background count rates for the \textit{XMM-Newton} MOS and PN cameras in the soft (0.4-2.0 keV), medium (2.0-5.0 keV), and hard (5.0-8.0 keV) bands. The colors and vertical dashed lines are the same as those displayed in Figure \ref{XrayLCMOS}.}\label{Back}
\end{figure*}
\end{center}

\vspace{-30pt} 

Figures \ref{XrayLCMOS} and \ref{XrayLCPN} show that the \textit{Swift} observations are consistent with the \textit{XMM-Newton} data in the soft band. In the total band, the \textit{Swift} data are consistent with the \textit{XMM-Newton} rates within uncertainties, except near phase 0.6, where the \textit{XMM-Newton} PN count rate is lower. The Swift data show less variability than the \textit{XMM-Newton} data due to their large phase binning. 

In the few places where the \textit{XMM-Newton} observations overlap in phase, they show good agreement and are clearly phase-locked, particularly in the soft band, even though the data sets were taken during different orbits of the system and were sometimes separated by several years. This is particularly evident near phases 0.0 and 0.5, where observations from different orbits and years would be indistinguishable from each other without the color coding in Figures \ref{XrayLCMOS} and \ref{XrayLCPN}. However, in the medium and hard bands, observations near primary eclipse (0818 and 2292) are not always consistent with each other. There also appears to be a fair amount of stochastic variability in the light curves, possibly due to instabilities in the system's stellar winds. For example, observations 2275 (dark blue in Figures \ref{XrayLCMOS} and \ref{XrayLCPN}), 0827 (light blue), and 0895 (yellow) exhibit apparently stochastic fluctuations in count rate around the global light curves. We find no correlation between these variations and the behavior of the background count rate (Figure \ref{Back}), which suggests they are intrinsic to the V444 Cyg system. However, these departures are similar in size to the uncertainties on the count rates, so their importance still remains unclear. Additional X-ray observations of V444 Cyg are needed to understand and determine the exact characteristics of this behavior. Next-generation X-ray observatories with much larger collecting areas will provide the precision needed to investigate these stochastic variations and their relationship to massive star winds. 

\subsection{\textit{XMM-Newton} spectra}

\begin{figure*}
\centering
\includegraphics[scale=0.29]{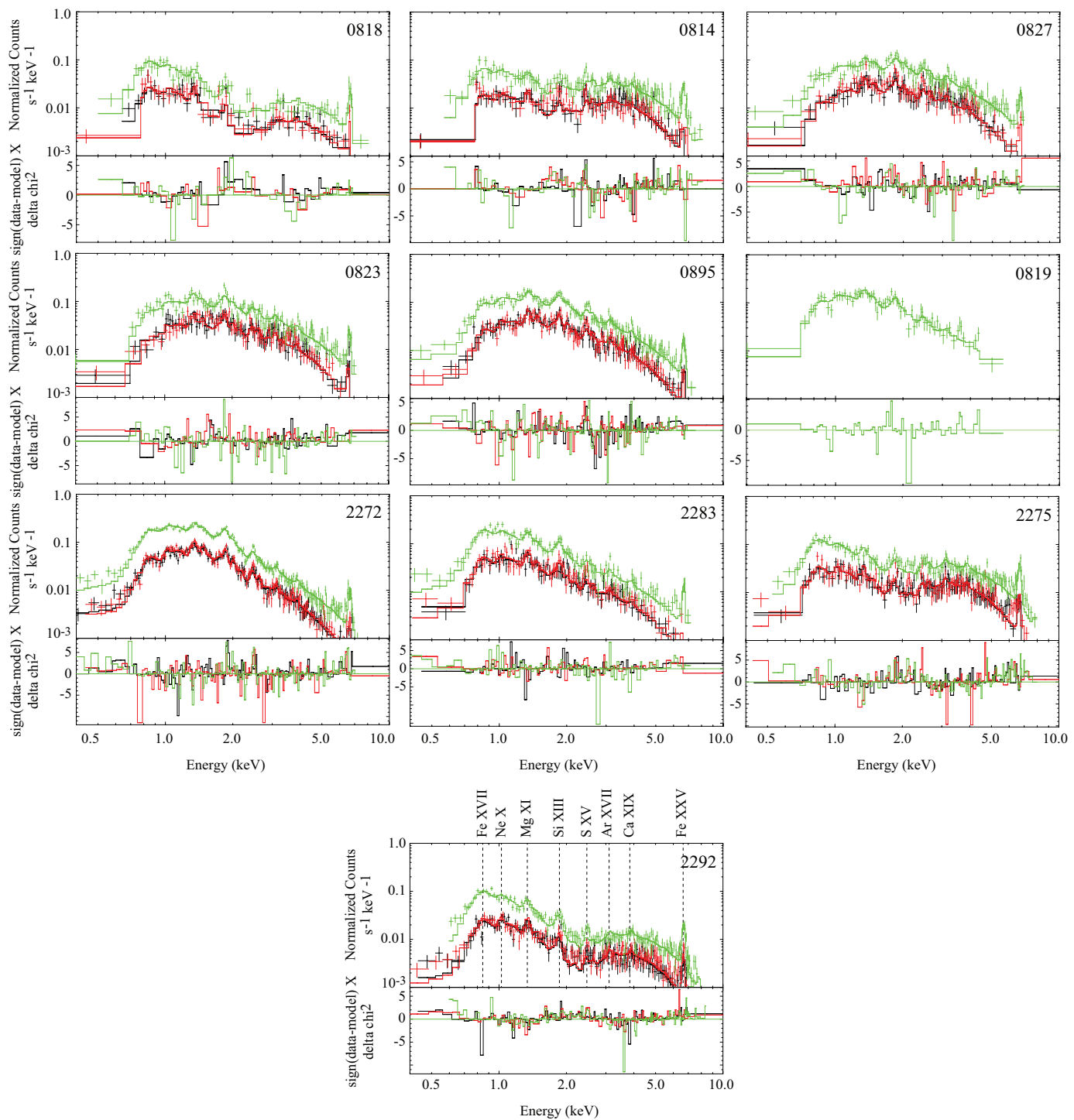}
\caption{\textit{XMM-Newton} EPIC spectra. Green (PN), red (MOS2), and black (MOS1) points represent observed data from the different detectors within the EPIC instrument. The solid lines in the same colors represent the XSPEC model fits discussed in Section 3.1. Spectra are indicated by their revolution number and arranged in phase order increasing from left to right across each row. Emission lines are identified in spectrum 2292.}\label{XraySpecSample}
\end{figure*}

We display all our extracted \textit{XMM-Newton} spectra in Figure \ref{XraySpecSample}. The four observations (revolution numbers 0814, 0818, 2275 and 2292) around primary eclipse exhibit a double-humped spectral shape, while all other spectra are single-peaked. We have identified many of the strong emission lines that appear in each observation and display those identifications in the last panel.

To interpret these spectra, we fit the data with the XSPEC (v12.7.1) software package using the following two-component model \citep{Arnaud} with the same binning as the extracted spectra
$$wabs \times (vphabs \times vapec + vphabs \times vapec), $$
where $vapec$ is an emission spectrum from a diffuse and collisionally ionized gas,  $vphabs$ is a photoelectric absorption component, and $wabs$ represents the interstellar medium absorption component with a hydrogen column density fixed to $n_H=0.32 \times 10^{22}$ cm$^{-2} $ \citep{Oskinova}. We refer to the abundance table of \cite{Anders} for the other components. 

For each observation we fit the three EPIC spectra simultaneously, except for observation 0206240401 (revolution 0819) where we only have PN data due to a strong flaring event. We performed the following careful step-by-step fitting procedure. First we allowed the temperatures, absorptions, and strengths (i.e. normalization factors within XSPEC) of the two components to vary freely, but we fixed the abundances to solar \citep{Anders}. We found the temperatures of the two components to be constant with phase within their uncertainties: 0.6 keV for the first component and 2.0 keV for the second component. This is consistent with the analysis by \cite{Maeda} and expected for a colliding wind in a circular orbit where preshock speeds are constant. Therefore, we froze the temperatures of each observation at those values.  

With these fixed temperatures we then allowed abundances to vary, linking the abundances of each individual element across the model components. This produced only one abundance per element during the fitting; in the following discussion we quote abundances in the XSPEC format (relative to hydrogen relative to solar). We allowed one abundance to vary at a time in the following order: nitrogen (N), silicon (Si), sulfur (S), neon (Ne), magnesium (Mg), carbon (C), oxygen (O), and iron (Fe). While varying each individual abundance, we held all the others constant. We found the abundance of N to be about three times solar and constant with phase within uncertainties (i.e. we do not detect WN abundances at some phases, and O-star abundances at others). Therefore, we froze it at that value before moving on to the next element. Similarly, we found the Si, S, and Ne abundances for all of the spectra were solar within uncertainties, so we froze them at 1.0. Magnesium was more abundant than solar; we froze it at 1.5. We found that C and O were comparable with a null value (likely due to high absorption providing little flux at C and O energies), so we set them to 0.0, while we found Fe to be just slightly lower than solar (0.8). These values are consistent with the results of \cite{Fauchez}, but differ significantly from the abundances found by \cite{Bhatt}, whose method of spectral modeling often resulted in unphysically large abundances. Figure \ref{XraySpecSample} shows the resulting model fit for each of our EPIC observations.

\begin{figure}
\includegraphics[scale=0.4]{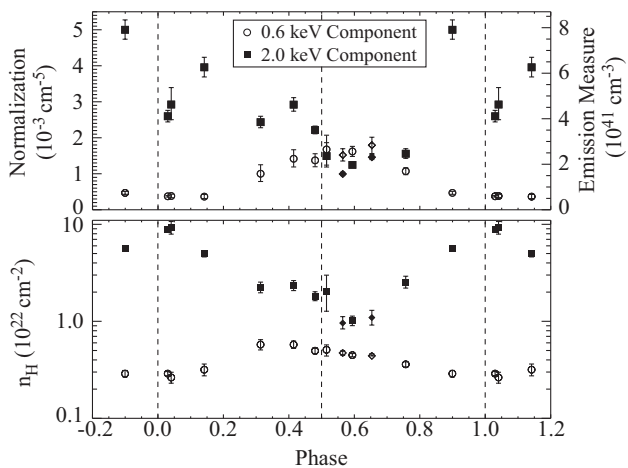}
\caption{Normalization (top) and absorption column density ($\textrm{n}_\textrm{H}$; bottom; y axis is log scale) parameters from our spectral fitting of the \textit{XMM-Newton} observations as functions of phase. Circles (0.6 keV component) and filled squares (2.0 keV component) represent the two different model components. In the case of observation 2272, we split the observation at phase 0.624 as well as fitting the whole observation (Section 3.1). Diamonds (open=0.6 keV, closed=2.0 keV) represent the two data sets derived from this split. Points are plotted in phase at the midpoint of each observation. Dotted vertical lines represent phases 0.0, 0.5, and 1.0. All data have been wrapped in phase so that more than one complete cycle is shown.}\label{XraySpecParam}
\end{figure}

After we completed the abundance fitting, the only free parameters remaining were the normalization and absorption columns (Table \ref{Abundance}). The stellar separation in V444 Cyg does not change as a function of orbital phase because the orbit is circular. Therefore, the intrinsic emission from the system should be constant with phase, barring orientation effects such as occultation and absorption by wind material. We allowed the normalization and column density to vary in order to test for these effects. We display our fits' resulting absorption columns and normalization parameters as functions of orbital phase in Figure \ref{XraySpecParam}. 

The absorption of the 0.6 keV component stays nearly constant over the cycle (column 5 of Table \ref{Abundance}; we note that the y axis in the bottom panel of Figure \ref{XraySpecParam} is logarithmic). The lack of eclipse effects suggests that the soft emission arises in the outer regions of the stellar winds of the two stars, especially the WN wind. This behavior is reminiscent of that of WR 6, a single WN5 star \citep{Oskinova12}. The soft X-ray emission in WR 6 originates so far out that conventional wind shock models do not provide an adequate understanding of the mechanisms that generate it. In V444 Cyg, the 0.6 keV absorption is not uniformly constant, however; its slight increase between phases 0.22 and 0.6 is likely due to the O-star wind's intrinsic emission becoming visible as the WN star and its wind rotate behind the O star. At these phases the WN wind is no longer absorbing the O-star emission along our line of sight. This increased absorption comes from the fact that the soft intrinsic emission of the O star is emitted close to the photosphere, where absorption is larger compared to WN stars where this emission occurs farther from the star. 

As expected, the absorption of the hotter component (2 keV) which arises from the wind-wind collision, is strongest when the WN star and its dense wind are in front of the wind collision region (phase 0.0). Similarly, the absorption of this component is weakest when the X-ray source is seen through the more tenuous O-star wind, between phases 0.25 and 0.75. The length of this phase interval (half the orbital period) suggests that the bow shock separating the O-star wind from the WN wind has a very large opening angle. We discuss this important result further in Section 4.

The absorption behavior of the spectral fits can be correlated with features in the soft X-ray light curve (Figures \ref{XrayLCMOS}, \ref{XrayLCPN}, and \ref{XraySpecParam}, and Section 3.1). The phases when the WN star and its wind are in front (near primary eclipse) are the phases with the strongest absorption; they are also the phases at which the soft count rate is the lowest. More of the soft X-rays can escape the system when the O star and its wind are in front because of that wind's lower absorption. The spectral fits show that the absorption of the 0.6 keV component remains relatively constant between phases 0.25 and 0.75, despite the large variation in the soft count rate, whereas the 2 keV absorption component is highest when the WN star is in front. This can be understood by considering the emission is arising from the O-star wind. When the WN star is in front, the intrinsic X-ray emission from the O-star wind is absorbed by the WN wind, but when the O star is in front its intrinsic emission becomes visible. This explains the increase in the normalization of the 0.6 keV component in the 0.25 to 0.75 phase interval (Figure \ref{XraySpecParam}) and the increase in the soft-band count rate (Figures \ref{XrayLCMOS} and \ref{XrayLCPN}). Moreover, we attribute the asymmetry of the soft light curve to distortion from Coriolis deflection of the shock cone. In this senario, the distortion happens in such a way as to shift its maximum visibility toward later phases (i.e. the peak is near phase 0.63 instead of centered around phase 0.5). We discuss the interplay between the emission and absorption behavior of the system further in Section 4.1.
  
In addition to the above spectral fitting, we divided observation 0692810401 (2272) into two separate spectra due to the large change in count rate over the course of the observation (see Figures \ref{XrayLCMOS} and \ref{XrayLCPN}). The observation was divided at phase 0.624 (2272\_1 before, 2272\_2 after) which is the approximate peak phase of the 0.4-2.0 keV light curve (Section 3.2).We performed spectral fitting on the two resulting spectra. We started our fitting process on the subdivided data with the best fit from the total observation (2272), keeping the same temperatures and abundances that were found to be constant on a global level (see previous discussion). The only free parameters were the normalization and absorption columns (Table \ref{Abundance} and Figure \ref{XraySpecParam}), which we found agreed with the total 2272 observation and the overall normalization and column density trends. 


\begin{center}
   \begin{table*}[t]
   \begin{center}
      \caption[]{Spectral Fit Information for V444 Cyg \textit{XMM-Newton} Observations}
         \label{Abundance}
   
        \begin{tabular}{llllllllll}
            \hline
            \hline
            Revolution & 0.4-2.0 keV & 2.0-5.0 keV &5.0-8.0 keV & Norm. 1st & n$_H$ 1st & Norm. 2nd & n$_H$ 2nd & Reduced & Degrees of\\
            & Flux & Flux & Flux &  &  &  &  & $\chi^2$ & Freedom\\
            \hline\\[-10pt]

    0814   &  $^{1.12E-13}_{2.30E-12}$ &  $^{5.97E-13}_{1.68E-12}$ &  $^{3.56E-13}_{4.07E-13}$  & $0.37_{0.05}^{0.06}$ & $0.32_{0.04}^{0.05}$ & $3.96_{0.27}^{0.28}$ & $4.97_{0.36}^{0.39}$ & 1.1498& 263\\[2pt]
        \hline\\[-10pt]
    0818   &  $^{1.28E-13}_{1.81E-12}$   & $^{2.53E-13}_{1.24E-12}$ & $^{2.35E-13}_{3.01E-13}$  & $0.38_{0.05}^{0.05}$ & $0.26_{0.03}^{0.03}$ & $2.92_{0.42}^{0.47}$ & $9.22_{1.30}^{1.47}$ & 1.0866 & 143\\[1pt] 
        \hline\\[-10pt]
    0819   &  $^{3.36E-13}_{2.33E-12}$ &  $^{4.62E-13}_{7.19E-13}$   &  $^{1.46E-13}_{1.54E-13}$  & $1.68_{0.44}^{0.40}$ & $0.51_{0.07}^{0.06}$ & $1.49_{0.30}^{0.37}$ & $2.04_{0.77}^{0.95}$ & 0.9666 & 66\\[1pt]
        \hline\\[-10pt]
    0823   &  $^{2.77E-13}_{2.79E-12}$ & $^{7.67E-13}_{1.31E-12}$   &  $^{2.83E-13}_{3.01E-13}$  & $1.42_{0.23}^{0.25}$ & $0.58_{0.05}^{0.05}$ & $2.92_{0.18}^{0.19}$ & $2.34_{0.26}^{0.29}$ & 1.1529 & 299\\[1pt]
        \hline\\[-10pt]
    0827   &  $^{2.08E-13}_{2.15E-12}$ &  $^{6.43E-13}_{1.08E-12}$  &  $^{3.36E-13}_{2.51E-13}$  & $1.00_{0.22}^{0.25}$ & $0.58_{0.07}^{0.07}$ & $2.43_{0.16}^{0.16}$ & $2.23_{0.27}^{0.30}$ & 1.1224 & 277\\[1pt]
        \hline\\[-10pt]
    0895   &  $^{2.08E-13}_{2.40E-12}$ &  $^{6.62E-13}_{1.01E-12}$   &  $^{2.13E-13}_{2.29E-13}$  & $1.37_{0.18}^{0.19}$ & $0.50_{0.03}^{0.03}$ & $2.22_{0.11}^{0.11}$ & $1.82_{0.18}^{0.20}$ & 1.1455 & 363\\[1pt]
        \hline\\[-10pt]
    2272  &  $^{4.37E-13}_{2.15E-12}$  &  $^{4.86E-13}_{6.19E-13}$  &  $^{1.26E-13}_{1.29E-13}$  & $1.62_{0.14}^{0.14}$ & $0.45_{0.02}^{0.02}$ & $1.25_{0.04}^{0.04}$ & $1.01_{0.11}^{0.12}$ & 1.2884 & 444\\[1pt]
        \hline\\[-10pt]
	2272\_1  &  $^{3.82E-13}_{1.94E-12}$ &  $^{4.03E-13}_{5.08E-13}$   &  $^{1.00E-13}_{1.03E-13}$  & $1.52_{0.17}^{0.18}$ & $0.47_{0.02}^{0.02}$ & $9.99_{0.05}^{0.05}$ & $0.96_{0.13}^{0.16}$ & 1.0173 & 333\\[1pt]
	 \hline\\[-10pt]
	2272\_2  &  $^{4.83E-13}_{2.43E-12}$ &  $^{5.55E-13}_{7.19E-13}$  &  $^{1.47E-13}_{1.51E-13}$  & $1.79_{0.23}^{0.22}$ & $0.44_{0.01}^{0.02}$ & $1.46_{0.07}^{0.07}$ & $1.09_{0.18}^{0.21}$ & 1.1711 & 388\\[1pt]        
 	 \hline\\[-10pt]          
    2275  &  $^{1.51E-13}_{2.92E-12}$ &  $^{6.65E-13}_{2.11E-12}$  &  $^{4.41E-13}_{5.15E-13}$  & $0.47_{0.04}^{0.04}$ & $0.29_{0.02}^{0.02}$ & $5.00_{0.26}^{0.27}$ & $5.70_{0.31}^{0.33}$ & 1.1992 & 379\\[1pt]
\hline\\[-10pt]
    2283  &  $^{2.98E-13}_{1.78E-12}$ & $^{4.14E-13}_{7.16E-13}$    &  $^{1.5E-13}_{1.61E-13}$  & $1.07_{0.10}^{0.10}$ & $0.36_{0.02}^{0.02}$ & $1.56_{0.13}^{0.14}$ & $2.54_{0.34}^{0.38}$ & 1.1940 & 278\\[1pt]
        \hline\\[-10pt]
        2292  &  $^{1.16E-13}_{1.64E-12}$  &  $^{2.36E-13}_{1.11E-12}$  &  $^{2.11E-13}_{2.68E-13}$  & $0.37_{0.02}^{0.02}$ & $0.29_{0.01}^{0.01}$ & $2.60_{0.16}^{0.16}$ & $8.86_{0.53}^{0.56}$ & 1.5771 & 418\\[1pt]
        \hline
         \end{tabular}\\
    \tablefoot{These values were calculated by spectral fitting with the XSPEC software package (Section 3.1). Fluxes (columns 2, 3 and 4), normalizations (columns 4 and 6), and column densities (columns 5 and 7) are in units of erg cm$^{-2}$ s$^{-1}$, cm$^{-5}$, and $10^{22}$ cm$^{-2}$ respectively. Absorbed (upper) and unabsorbed (lower) fluxes are given for each observation in the soft, medium, and hard bands for which light curves were created (Section 3.2). Normalizations and column densities have high and low error bars as indicated. In the case of revolution 2272, we split the observation at phase 0.624 and performed spectral fitting of the resulting spectra as well as the whole observation (Section 3.1). Observation 2272\_1 represents the spectra derived from the data between the start of observation 2272 and phase 0.624, while 2272\_2 is from phase 0.624 to the end of the observation.  See Section 3.1 for abundances used for all fits.}
    \end{center}
   \end{table*} 
\end{center}   

\subsection{Optical polarimetry}

\begin{figure}
\includegraphics[scale=0.5]{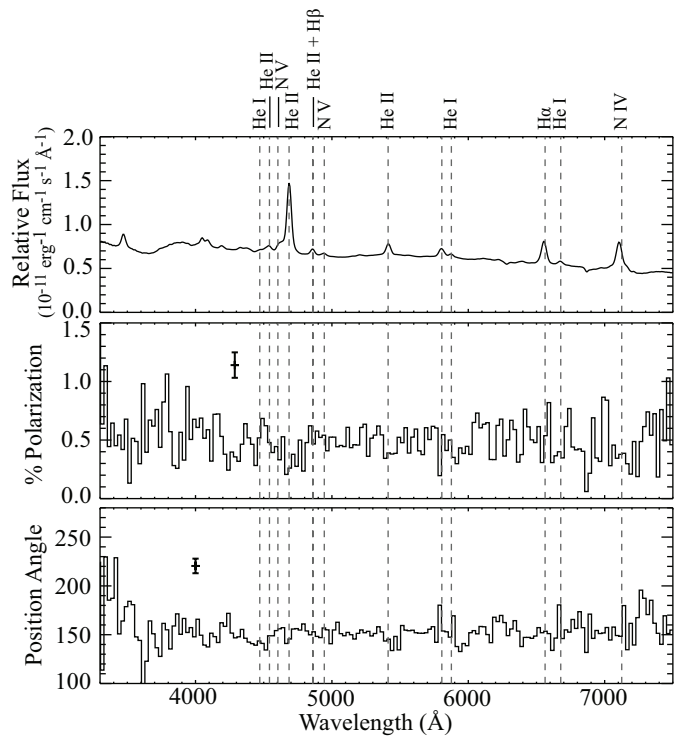}
\caption{Error-weighted mean polarization spectrum for HPOL@PBO observations between phases 0.6 and 0.75. \textit{From top:} relative flux, percent polarization, and position angle (degrees) versus wavelength. Gray dashed lines indicate identified emission lines. The polarization and position angle data have been binned to 25\AA\space. Error bars shown are the average polarization and position angle errors for the spectrum.}\label{PolSpec}
\end{figure}

Electron scattering in the ionized WN wind of V444 Cyg causes the system's observed polarization \citep{Robert,StLouis}. Continuum polarization light curves show phase-dependent variations that cannot be completely described by the standard, sinusoidal, `BME'-type behavior expected from detached binary stars \citep{BME,StLouis}. Deviations from that behavior near secondary minimum, including an asymmetric eclipse, indicate that the geometry of the WN wind is aspherical, likely due to the orbital motion of the system \citep{StLouis}. The emission lines in V444 Cyg also posses both linear and circular polarization; \cite{Chevrotiere} analyzed circular polarization measurements of the He II $\lambda$ 4686 emission line and detected no magnetic field. These authors noted the presence of variable linear polarization in the line, but attributed it to binary orbital effects. However, because emission line photons may scatter in different locations within the wind than continuum photons, detailed analyis of linear line polarization variations can reveal new information about the wind structure in V444 Cyg. We took advantage of the spectropolarimetric nature of our HPOL data to investigate the polarization behavior of the strongest emission lines in the optical regime. Figure \ref{PolSpec} shows a sample polarization spectrum produced by taking the error-weighted mean of the four HPOL@PBO observations between phases 0.6 and 0.75 (Table \ref{HPOLInfo}), and binning the resulting stacked spectrum to 25\AA; we have labeled many of the major emission lines.

Our observations contain polarization due to scattering by interstellar dust as well as electron scattering within the V444 Cyg system. In a line region, the polarization produced within the system may include separate components for continuum and line polarization if they are scattered in different regions of the winds. Therefore, the observed polarization at each wavelength within an emission line is a vector sum of the three effects
\begin{equation}\hspace{20pt}\label{obsq} \%q_{\text{obs}}=\frac{Q_{\text{c}}F_{\text{c}}+Q_{\text{L}}F_{\text{L}}+Q_{\text{ISP}}F_{\text{tot}}}{F_{\text{tot}}}, 
\end{equation} 
where $Q_\text{c}$, $Q_\text{L}$, and $Q_\text{ISP}$ represent the \textit{Q} Stokes parameters describing the continuum, line, and interstellar polarization, respectively, while $F_\text{c}$ and $F_\text{L}$ denote the flux in the continuum and line with $F_\text{tot}=F_\text{c}+F_\text{L}$. Parallel versions of this equation and those below also hold for the Stokes \textit{U} parameter. We note that because these are vector equations, the sign of each \textit{Q} (\textit{U}) parameter may be either positive or negative.

We calculated the polarization in the HeII $\lambda$4686, H$\alpha$, and NIV $\lambda$7125 emission lines using the flux equivalent width (few) method described by Hoffman et al. (1998). This allows us to separate the polarization into its continuum and line components without making any assumptions about the system. With this method we estimate the continuum flux and polarized flux within the line regions and subtract these quantities from the numerator and denominator of Equation \ref{obsq}
\begin{equation}\hspace{10pt}
 \%q_{\text{few}}=\frac{Q_{\text{c}}F_{\text{c}}+Q_{\text{L}}F_{\text{L}}+Q_{\text{ISP}}F_{\text{tot}}-(Q_{\text{c}}F_{\text{c}}+Q_{\text{ISP}}F_{\text{c}})}{F_{\text{tot}}-F_{\text{c}}}.
\end{equation} 
This leaves, for each wavelength in the line region,
\begin{equation}\hspace{20pt}\label{finalpol}
 \%q_{\text{few}}=\frac{Q_\text{L}F_\text{L}+Q_\text{ISP}F_\text{L}}{F_\text{L}}=\%q_\text{L}+\%q_\text{ISP}.
\end{equation}
To obtain a total polarization value for the line, we sum both numerator and denominator of Equation \ref{finalpol} over all contributing wavelengths, which we choose as described below. Because the interstellar polarization varies slowly with wavelength \citep{Serkowski,Whittet}, $Q_{\text{ISP}}$ is very close to constant over the wavelength region of an emission line. Thus the total polarization within an emission line is given by

\begin{equation} 
\begin{aligned}
\%q_{\text{entire line}} &= \frac{(Q_\text{L,1}F_\text{L,1}+Q_\text{ISP}F_\text{L,1})+(Q_\text{L,2}F_\text{L,2}+Q_\text{ISP}F_\text{L,2})+...}{F_\text{L,1}+F_\text{L,2}+...}\\
&=\frac{Q_\text{L,1}F_\text{L,1}+Q_\text{L,2}F_\text{L,2}+...}{F_\text{L,1}+F_\text{L,2}+...}+\frac{Q_{\text{ISP}}(F_\text{L,1}+F_\text{L,2}+...)}{F_\text{L,1}+F_\text{L,2}+...}\\
&=\% q_\text{line all}+ \% q_\text{ISP}.
\end{aligned}
\end{equation}

In our discussion below, we report our measurements of $\% q_\text{entire line}$ and $\% u_\text{entire line}$, each of which is a simple vector sum of the interstellar polarization and polarization produced within the system. We have made no attempt to remove ISP effects from our line polarization data, so these data cannot be taken to represent absolute line polarization values. However, the constant ISP contribution produces only a simple linear shift in the zero point of the polarization in Figures 8-11. Because we expect that any ISP contribution will not vary with time over the period of our observations, we can attribute any time variability in the results to changes in polarization arising within the V444 Cyg system.

\begin{figure}
\includegraphics[scale=0.5]{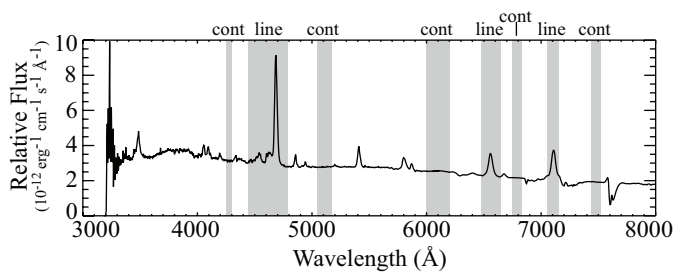}
\caption{Relative total flux spectrum of V444 Cyg from 2012 Oct 22 using the HPOL@Ritter instrument setup. Gray areas mark the line and continuum regions we used to calculated the line polarization with the flux equivalent width method (Section 3.3) for each observation. We determined the underlying continuum for each line using the shaded continuum regions immediately to its red and blue sides. }\label{Regions}
\end{figure}

In order to estimate the continuum flux and polarization for the subtraction in Equation 2, we fit linear functions to the flux and polarization spectra between two wavelength regions on either side of the emission line (Figure 7). The choice of regions from which to estimate the continuum is important because the ISP has a shallow wavelength dependence, and because the inclusion of a polarized line in a continuum region will skew the final polarization calculated for the line of interest. Therefore, we chose continuum regions near the emission line of interest that appear to have no emission or absorption features in our total or polarized spectra. Figure \ref{Regions} shows the continuum and line regions we used in our calculations. 

In the case of the HeII $\lambda$4686 line we chose a line region that includes not only this line but also the NV $\lambda$4604 4620 doublet and the HeII $\lambda$4542 and HeI $\lambda$4471 lines, which are blended with each other in our spectra. This makes it more difficult to determine which line is responsible for any phase-dependent polarization behavior; however, this ensures that we include all spectral regions which may be contributing to any variations. Similarly, the NIV $\lambda$7125 line has many absorption and emission lines around it, but we were able to define appropriate continuum regions and a line region that includes the line core without other stray lines. In the case of the H$\alpha$ line we were able to define line and continuum regions more easily than for the other lines. 

Photospheric absorption in the O star is an important source of line profile variations observed in total light \citep{Marchenko,Flores}. If we do not correct for this unpolarized absorption we risk removing too much continuum polarization from our observations. We can correct for underlying absorption features by treating them as part of the continuum in Equation 2; this reduces the value of $F_\text{c}$ so as to produce a specified absorption equivalent width, as detailed by \cite{HNF}. \cite{Marchenko} removed an absorption component from their V444 Cyg HeII $\lambda$4686 profile data by approximating the O-star absorption as a Gaussian line profile with a full width at half maximum (FWHM) of 6.0 \AA\space and an EW of 1.0 \AA. The same authors estimated an absorption profile for the HeI $\lambda$4471 line as a Gaussian with a FWHM of 6.0\AA\space and an EW of 0.25 \AA. While we do not know the EW of the NV doublet that is also blended with the HeII $\lambda$4686 line, we assumed an underlying unpolarized absorption component using a total EW of 2 \AA\space. This value corrects for the absorption due to the He lines (1.25 \AA\space total) while estimating a total absorption equivalent width of 0.75 \AA\space for the N doublet. We found that the corrected line polarization values were not significantly different from the uncorrected values. Similarly, we found that for the H$\alpha$ line, corrections on the order of half the total EW need to be made before the corrected values differ from the uncorrected by more than their uncertainties. This is significantly larger than the absorption EWs estimated for O stars \citep[2\AA]{Kurucz}. Therefore, in the rest of this paper we present only the uncorrected data for all lines.

Table 6 tabulates the line polarization results we obtained using the methods detailed above. To interpret these data, we need to understand how polarization is formed in the V444 Cyg system. Because both stars in the system are hot, we expect that electron scattering is the dominant polarizing mechanism for both line and continuum light. Zeeman splitting of spectral lines due to magnetic fields can give rise to second-order effects in the Stokes \textit{Q} and \textit{U} parameters \citep{Petit}. However, a recent study of V444 Cyg designed to detect Zeeman features in circular polarization (Stokes \textit{V}) did not detect a magnetic field in the system \citep{Chevrotiere}, so this effect is unlikely to cause the linear polarization we observe.

Even if electron scattering is the only polarizing mechanism at work, light in the emission lines may form, scatter, and/or become eclipsed in different regions than does light in the continuum. Analyzing these variations can help us further constrain the geometry of the emission and scattering regions in V444 Cyg. Although our data include a constant ISP contribution as discussed above, their behavior with time provides important clues to the geometry of the line-scattering regions within V444 Cyg. Any variability in the line polarization indicates that the geometry of the system's line-scattering region(s) is changing over time, due either to our changing perspective as the system rotates or to intrinsic changes in the distribution of scatters. If we observe no line polarization variability, we can assume that the geometry of the relevant line-scattering region remains constant over the time period of our observations.

\begin{figure}
\includegraphics[scale=0.5]{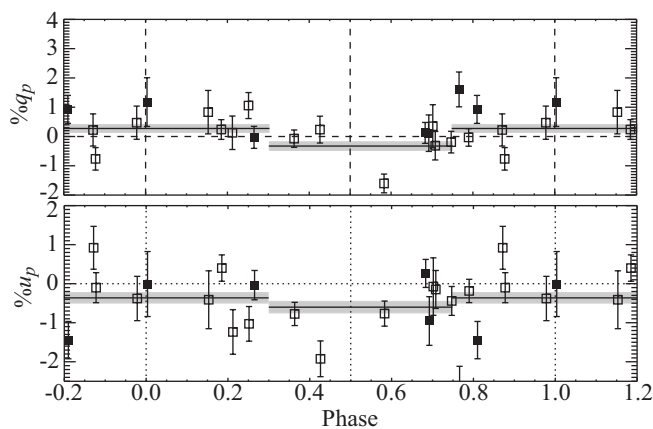}
\caption{Data points represent the HeII $\lambda$4686 emission line polarization from HPOL@PBO (open squares) and HPOL@Ritter (filled squares). From top: projected \% \textit{q}$_p$ Stokes parameter and projected \% \textit{u}$_p$ Stokes parameter versus phase (see text; rotated by $-15\degree$). The solid horizontal lines mark error-weighted mean values of \% \textit{q}$_p$ and \% \textit{u}$_p$ for the phase regions they span; 1$\sigma$ uncertainties are shown in gray. Dotted vertical lines represent phases 0.0, 0.5, and 1.0. Dotted horizontal lines mark zero in \% \textit{q}$_p$ and \% \textit{u}$_p$.}\label{HeFigure}
\end{figure}

\begin{figure}
\includegraphics[scale=0.5]{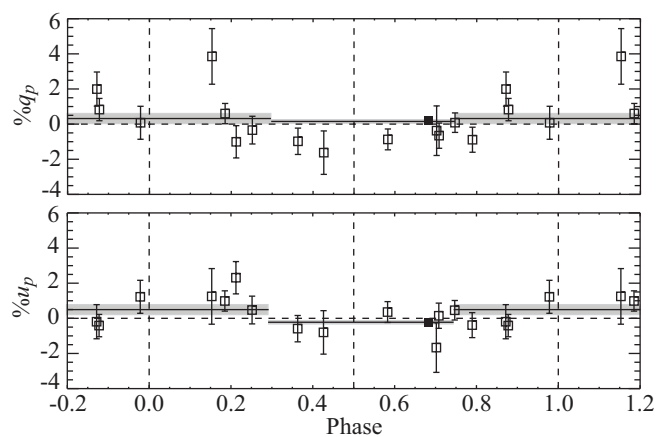}
\caption{Same as Figure \ref{HeFigure}, but for the NIV $\lambda$7125 line. Stokes parameters are rotated by $10.4\degree$.}\label{NFigure}
\end{figure}

\begin{figure}
\includegraphics[scale=0.5]{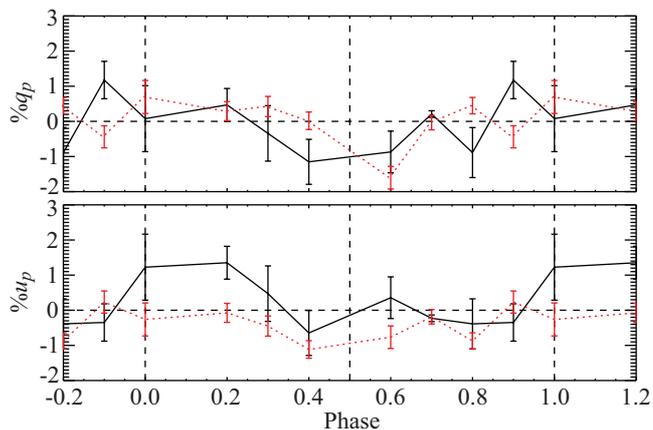}
\caption{Unrotated Stokes parameters for the HeII $\lambda$4686 (red dotted) and NIV $\lambda$7125 (black solid) line data, binned to 0.1 in phase.}\label{many}
\end{figure}

\begin{figure}
\includegraphics[scale=0.5]{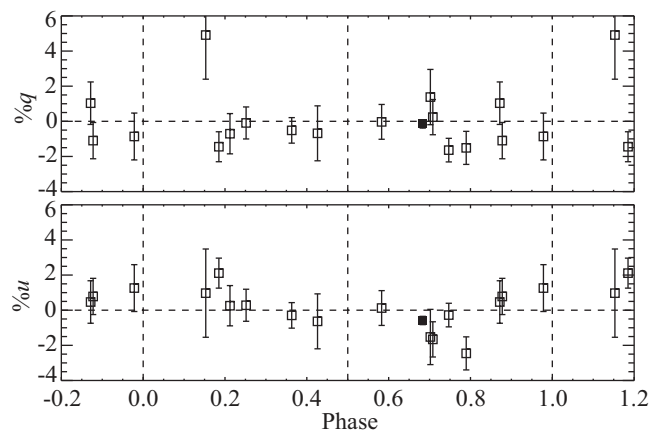}
\caption{Same as Figure \ref{HeFigure}, but for the H$\alpha$ line. Stokes parameters are unrotated.}\label{HaFigure}
\end{figure}

Figures \ref{HeFigure}-\ref{HaFigure} show the phase-dependent polarization behavior for the HeII $\lambda$4686, NIV $\lambda$7125, and H$\alpha$ lines (Table \ref{HPOL}). While emission lines in WR spectra are often unpolarized \citep{Harries}, our results show that this is not the case for V444 Cyg. If the lines were intrinsically unpolarized, then they would display a constant polarization due to the ISP contribution, with no phase-dependent behavior. However, Figures 8-10 show that the polarization in the HeII $\lambda$4686 and NIV $\lambda$7125 lines varies with phase, implying that these lines contain polarization contributions arising from scattering within the system. By contrast, the H$\alpha$ line polarization is consistent with zero in nearly all our observations. This could indicate either that it is intrinsically unpolarized (in which case the ISP contribution is necessarily small), or that its polarization is roughly constant with phase and nearly cancels out the ISP contribution. We discuss this line behavior further at the end of this section.

Figures 8-11 also show that the polarization behavior of these three lines is not consistent with the broadband polarimetric behavior found by \cite{StLouis}, which shows the standard sinusoidal behavior expected from binary stars along with an asymmetric secondary eclipse effect due to distortion of the WN's wind. This indicates that the continuum and line photons are polarized by different scattering regions within the system.

   \begin{table*}[t]
   \begin{center}
      \caption[]{HPOL Line Polarization Stokes Measurements}
         \label{HPOL}
   
        \begin{tabular}{lrrrrrrrrrr}
            \hline
            \hline
           
           & & \multicolumn{3}{c}{H$\alpha$} & \multicolumn{3}{c}{HeII $\lambda$4686} & \multicolumn{3}{c}{NIV $\lambda$7125}\\
            Date & Phase & \% \textit{q} & \% \textit{u} & \% Error & \% \textit{q} & \% \textit{u} & \% Error & \% \textit{q} & \% \textit{u} & \% Error \\
            \hline

    1989 Oct 03 & 0.790 & -1.5110 & -2.4579 & 0.9408 & 0.0654 & -0.1774 & 0.2992 & -0.6922 & -0.6771 & 0.7119 \\
    1990 Jul 24 & 0.583 & -0.0306 & 0.1236 & 0.9911 & -1.0061 & -1.4650 & 0.3212 & -0.9406 & 0.0248 & 0.5935 \\
    1990 Oct 23 & 0.185 & -1.4468 & 2.1103 & 0.8538 & 0.0068 & 0.4654 & 0.3375 & 0.2097 & 1.1375 & 0.5768 \\
    1990 Oct 24 & 0.363 & -0.5154 & -0.2944 & 0.7267 & 0.3231 & -0.7084 & 0.2973 & -0.7041 & -0.8975 & 0.7494 \\
    1991 Aug 21 & 0.878 & -1.0984 & 0.7828 & 1.0322 & -0.6105 & -0.4702 & 0.3845 & 0.9202 & -0.0931 & 0.6354 \\
    1991 Aug 31 & 0.252 & -0.0926 & 0.2813 & 0.9119 & 1.4296 & -0.3627 & 0.4441 & -0.4877 & 0.3206 & 0.7898 \\
    1991 Oct 10 & 0.747 & -1.6390 & -0.2771 & 0.6717 & 0.0549 & -0.4787 & 0.3698 & -0.0859 & 0.4593 & 0.5554 \\
    1991 Nov 17 & 0.708 & 0.2388 & -1.6601 & 1.0027 & -0.1982 & -0.2855 & 0.4866 & -0.6661 & -0.0959 & 0.7179 \\
    1994 Jun 15 & 0.153 & 4.9108 & 0.9717 & 2.5127 & 0.9262 & 0.0591 & 0.7414 & 3.1580 & 2.5385 & 1.5831 \\
    1994 Jul 03 &0.426 & -0.6828 & -0.6338 & 1.5643 & 1.1650 & -1.5495 & 0.4584 & -1.2326 & -1.3260 & 1.2366 \\
    1994 Aug 16 & 0.872 & 1.0321 & 0.4674 & 1.2103 & -0.2664 & 0.9080 & 0.5495 & 1.9333 & 0.5301 & 0.9724 \\
    1994 Oct 06 & 0.978 & -0.8630 & 1.2595 & 1.3337 & 0.5983 & -0.0933 & 0.5676 & -0.3639 & 1.1728 & 0.9390 \\
    1994 Nov 21 & 0.702 & 1.3809 & -1.5241 & 1.5728 & 0.3399 & 0.1109 & 0.7363 & 0.2433 & -1.6902 & 1.4111 \\
    1994 Dec 01 &0.212 & -0.7050 & 0.2555 & 1.1458 & 0.7282 & -1.0072 & 0.5716 & -1.7630 & 1.8021 & 0.9190 \\
    2012 May 11 & 0.810 & ... & ... & ... & 1.5238 & -0.7899 & 0.4783 & ... & ... & ... \\
    2012 Jul 13 & 0.766 & ... & ... & ... & 2.7477 & -1.5460 & 0.5949 & ... & ... & ... \\
    2012 July 14 & 0.004 & ... & ... & ... & 1.0221 & 0.5791 & 0.8323 & ... & ... & ... \\
    2012 Oct 22 & 0.683 & -0.1479 & -0.5860 & -0.0255 & 0.2886 & 0.3585 & 0.3585 & 0.3044 & -0.1509 & 0.0972 \\
    2012 Oct 26 & 0.692 & ... & ... & ... & 0.5757 & -0.7716 & 0.6236 & ... & ... & ... \\
    2012 Dec 14 & 0.265 & ... & ... & ... & -0.0076 & -0.0467 & 0.3757 & ... & ... & ... \\
    \hline
\end{tabular}\\
\end{center}
   \end{table*} 

Since the HeII $\lambda$4686 emission line shows little phase dependence in the Stokes \textit{Q} parameter, we calculated the error-weighted angle ($-15\degree \pm 5.8\degree$ or equivalently $165\degree \pm 5.8\degree$) by which we would have to rotate our data to place all of the polarization variation in the resulting projected Stokes \textit{U} parameter (Stokes \textit{U} marks the $45\degree$-$135\degree$ axis in unprojected \textit{Q-U} space). This rotation causes the projected Stokes \textit{Q} (which we depict in Figure 8 as a percentage, \% \textit{q}$_p$) to average to zero. We do not interpret this rotation angle as an intrinsic axis within the system; rather, it is a tool to simplify the data for display and aid in our interpretation. In the rest of this section, we discuss the polarization behavior of the HeII $\lambda$4686 only line in terms of \%\textit{u}$_p$. We rotated the NIV $\lambda$7125 line by $10.4\degree$ for the same reasons (uncertainty on the error-weighted mean is $\pm 11\degree$), with the same caveat that this angle is not necessarily intrinsic to the system. In the case of the H$\alpha$ line, both Stokes parameters remain zero within uncertainties for most of our observations. We therefore did not rotate these data.

After rotation, the values of \% \textit{q}$_p$ for the HeII $\lambda$4686 line remain positive for the first half of the light curve, but scatters around zero more evenly during the second half (Figure \ref{HeFigure}). In \% \textit{u}$_p$, the polarization shows variations of the same magnitude (on the order of $\pm 2$\%). Points between phases 0.6 and 1.0 have a larger scatter than the rest of the light curve, while observations near secondary eclipse have a lower \% \textit{u}$_p$ than the rest of the light curve. This lower pattern suggests that suggests V444 Cyg has a phase-locked polarization behavior. To quantify this behavior, we calculated the error-weighted mean \% \textit{u}$_p$ for the phase regions 0.30 to 0.75, and 0.75 to 1.30. Our choice of these two phase regions was guided by the differing behavior of the normalization and column densities seen in the X-ray spectra around secondary eclipse (Figure \ref{XraySpecParam}). The region around secondary eclipse (phases 0.3 to 0.75) that has a low 2 keV absorption in X-rays also has a lower average \% \textit{u}$_p$ ($-0.703\pm0.142$) than the rest of the light curve ($-0.193\pm0.133$), a discrepancy of nearly 2$\sigma$ (compare Figures \ref{XraySpecParam} and \ref{HeFigure}). The \% \textit{q}$_p$ averages for those same phase intervals overlap within uncertainties ($0.056\pm0.142$ for phases 0.30-0.75; $0.320\pm0.133$ for phases 0.75 to 1.30).

The NIV $\lambda$7125 line behaves similarly to the HeII $\lambda$4686 line (Figure \ref{NFigure}); the \% \textit{u}$_p$ values for this line are predominantly negative around secondary eclipse (phases 0.3-0.75) where the X-rays have a low 2 keV component absorption feature (Figure \ref{XraySpecParam}) and are positive or zero at other phases. However, the \% \textit{q}$_p$ values are negative in the 0.3 to 0.6 phase range, but scatter equally about zero at other phases. A similar analysis shows the error-weighted mean \% \textit{u}$_p$ for observations between 0.3 and 0.75 in phase is significantly lower ($-0.139\pm 0.085$) than the rest of the cycle ($0.579 \pm 0.279$), while the average \% \textit{q}$_p$ values for these two regions overlap ($0.228\pm0.085$ for phases 0.30-0.75; $0.110\pm0.279$ for phases 0.75 to 1.30).

To directly compare the behavior of the two lines, we smoothed our polarization data by calculating the error-weighted mean unrotated \%\textit{q} and \%\textit{u} for bins of width 0.1 in phase; when we overplotted the smoothed curves (Figure \ref{many}), we found that these two lines show similar phase-dependent behavior. The \%\textit{u} values are more positive near phase 0.0 and first quadrature (phase 0.25) than around phase 0.5 while after secondary eclipse they gradually trend toward more positive values. In contrast, \%\textit{q} remains relatively flat during the first half of the orbit, while the variations in the second half are of a more stochastic nature. 

We interpret these line polarization results as follows. Within the V444 Cyg system, the NIV emission should arise within the wind of the WN star, and the HeII is largely in a shell of material around the WN star (see Figure 8 in Marchenko et al. 1997). If these lines arose in the shells of HeII $\lambda$4686 and NIV $\lambda$7125 around the WN star and scatter in the spherically symmetric WN wind, then we would measure no net polarization. The fact that we measure a phase-dependent polarization behavior in both lines suggests that the wind must be aspherical in some way. The WN wind is much more dense than the O-star wind \citep{Stevens,Hirv}, so one simple way to visualize this asymmetry is to posit that the O-star wind carves out a less dense cavity within the WN wind. \cite{Kurosawa} used such a sphere plus cavity model to reproduce the observed continuum polarization variations in V444 Cyg. The emission line polarization should also contain signatures of this asphericity, but these signatures will differ from those seen in the continuum polarization because the line photons originate from the WN wind instead of form the stellar photospheres and may scatter in different locations within the wind.

For a sphere plus cavity model better suited to interpret our line polarization results, we turned to \cite{Kasen}, who investigated the polarization behavior of a hole in the ejecta of a Type Ia supernova due to its unexploded companion star. While the density profiles in this scenario are different than those of the winds in V444 Cyg, this model is highly relevant to our case for several reasons. First, the ejecta hole produces a cone-shaped cavity resembling the shock-wind structure within V444 Cyg. Second, in the Kasen et al model, the density of material within the hole is 0.05 times that of the ejecta, which is analogous to the less dense O-star wind filling the cavity within the denser WN wind. (We note that the wind density ratio in V444 Cyg is difficult to calculate because it depends on the wind velocities, which are in turn affected by the radiative forces discussed in Section 4.1). Third and most importantly, \cite{Kasen} considered a distributed emission source, with photon packets arising from throughout their model atmosphere to simulate the energy deposition from decaying $^{56}$Ni and $^{56}$Co into the supernova ejecta. This distributed emission is also a good representation of the line-emitting shells surrounding the WN star in V444 Cyg \citep{Marchenko}.

\cite{Kasen} oriented their models such that all the polarization variations occurred in one Stokes parameter. They found that for viewing angles near the axis of the hole, the polarization is negative, but it becomes more positive as the viewing angle moves away from the hole. Because we rotated our line data (Figures \ref{HeFigure} and \ref{NFigure}) such that most of the polarization variation occurred in \% \textit{u}$_p$, we can draw an analogy between these variations and those predicted by the ejecta-hole model. Using this analogy, we can attribute the change in polarization behavior of our HeII $\lambda$4686 and NIV $\lambda$7125 lines to a `hole' created by the shock cone (which we detect in our X-ray data) in the otherwise spherical shells around the WN star. As our viewing angle changes with phase, the changing geometry of the incomplete shells causes the variations in polarization we observe. In this picture, the `hole' is open to our light of sight at phases near secondary eclipse. The angle by which we rotated each line's data thus corresponds to the geometrical offset between the orientation of the shells in V444 Cyg and the orientation of the \cite{Kasen} model. These angels are similar for the two lines, but not the same; in addition, the lines show different polarization behavior with phase (Figure \ref{many}). If the ejecta-hole model is a good approximation of the winds in V444 Cyg, these discrepancies may simply indicate that the shock is a more complex structure than a cone and that the shells of material where lines form are misaligned. 

Unsurprisingly, our data do not exactly reproduce the trends in polarization with viewing angle (phase) predicted by \cite{Kasen}. One important difference between the case of V444 Cyg and the ejecta-hole scenario is the size of the opening angle of the shock hole. Our shock likely has a very large opening angle (see Sections 3.1 and 4.2), which means our viewing angle always remains closer to the cavity than in Kasen's models. We also have a shock structure with the same geometric shape as the hole, which makes it difficult to distinguish between light scattered in the wind and light scattered in the shock. In addition, although we rotated our data in an attempt to confine the phase-locked polarization to the \textit{U} Stokes parameter, both emission lines still display relatively large changes in \% \textit{q}$_p$ (Figures \ref{HeFigure} and \ref{NFigure}). In the context of the cavity model, this might indicate that at phases when we do not see the shock and the hole, the WN wind appears elongated rather than spherically shaped. 

In contrast to the HeII $\lambda$4686 and NIV $\lambda$7125 lines, the H$\alpha$ emission in V444 Cyg likely comes from cooler regions around the system. Figure \ref{HaFigure} shows that both Stokes parameters in the H$\alpha$ line are zero within uncertainties in the majority of our observations. This implies that the polarization in this line is constant with phase, although it may not have an intrinsic value of zero because we have not removed the ISP contribution. A constant polarization indicates that the scattering region for the H$\alpha$ line is far enough from the stars that their orbital motion not affect its geometry. The deviations from zero measured polarization do not appear to have a phase-dependent behavior, which suggests that clumping of material within the cooler regions of the system may be responsible for the 
variability \citep{PolClumps}. Such clumps are likely transient and unconnected to the orbital period of the system; thus, any polarization produced by scattering in this clumpy wind should be stochastic as suggested in Figure \ref{HaFigure}.

The emission lines in V444 Cyg are polarized at least in part by scattering within the system; they show polarization variations different from those of the continuum because the line and continuum photons originate from different regions. Despite these distinctions, our spectropolarimetric results support the cavity model developed by \cite{Kurosawa} to describe the wind structure in the system. The phase correlation we observe between the decrease in 2 keV X-ray column density (Section 3.2) and the decrease in emission line polarization provides the first concrete evidence tying polarimetric variations to the presence of a low-density cavity in the WN wind.

\section{Discussion} 

The behavior we see in both the X-rays and spectropolarimetry can be explained by a combination of three effects: 1) simple geometric eclipses; 2) distortion of the stellar winds due to the orbital motion of the stars; and 3) the presence of a cavity in the WN wind produced by its collision with the wind of the O star. Our X-ray light curves and spectra provide direct constraints on the geometry of the interacting winds in the system. Below we discuss simple models of these data that allow us to draw some basic conclusions about the V444 Cyg system. We adopt the system parameters in Table \ref{SysParam} and a separation of $a=35.97\;R_\sun$ \citep{Eris}. We do not calculate formal uncertainties on our models' outputs because of their simplicity; it is clear a more sophisticated treatment of the system is needed, which we plan to address in a future paper.

\subsection{Modeling the X-ray Light Curves}

\begin{figure*}[t]
\centering
  \leftline{\hspace{-160pt}
  \includegraphics[scale=0.7]{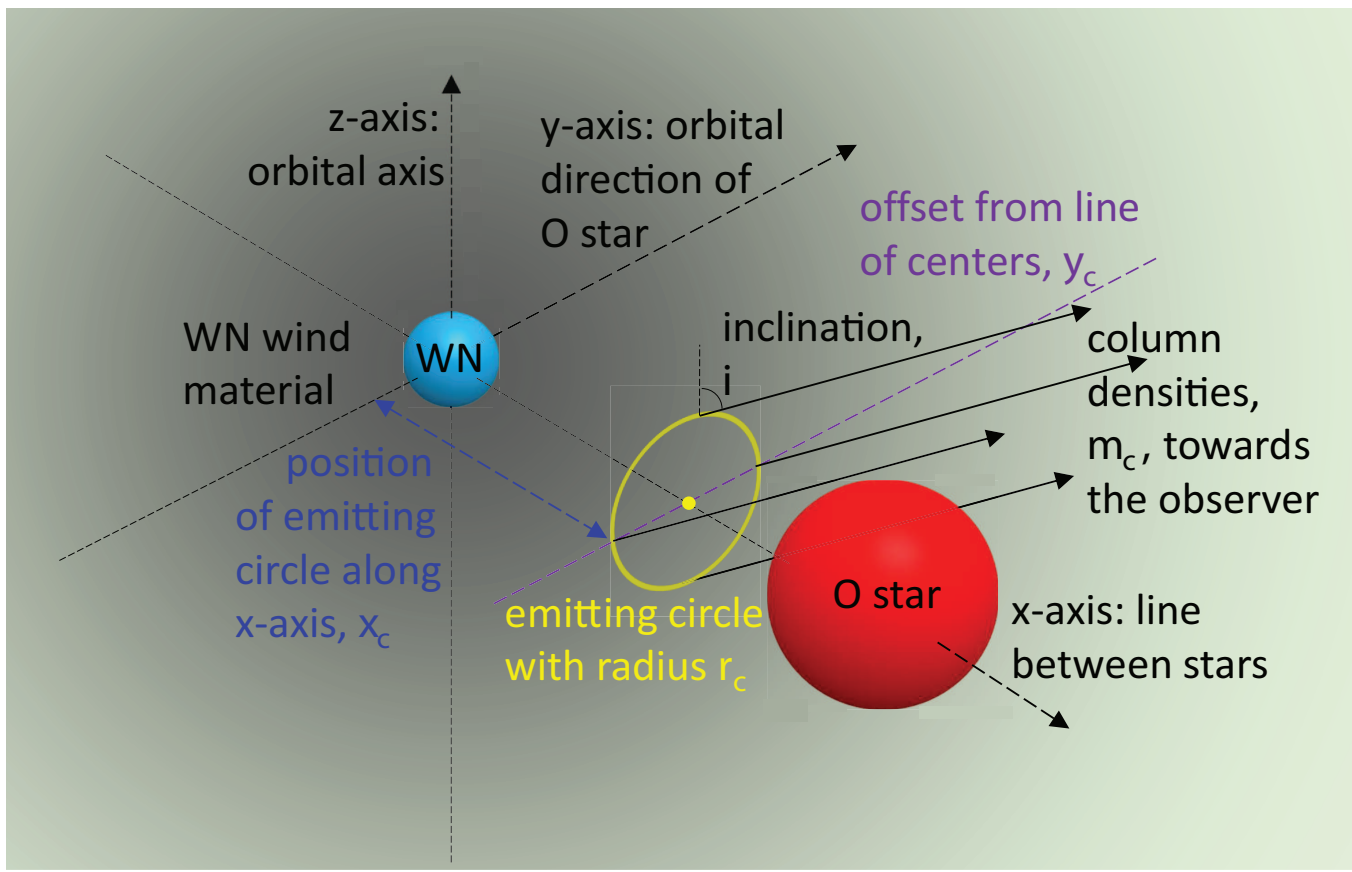}}
  \caption{Schematic of the occultation plus WR-wind--absorption model showing four column density calculations from the circle of emission to the observer (specifically from the most pro/retrograde portions of the circle and above/below the line of centers).  The observer is rotated around the system in the $xy$-plane to compute the relative phase-dependent variations of the light curve.}
  \label{fi:ChrisModel}
\end{figure*}

To further interpret the phase variation of the {\it XMM} data in the soft, medium, and hard bands, we modeled the colliding-wind X-rays of V444 Cyg with an analytic occultation plus WN-wind--absorption model with a geometry shown in Figure \ref{fi:ChrisModel}. We define the $x$-axis as the line of centers between the stars; our modeled X-ray emission then originates from a circle (yellow in Figure \ref{fi:ChrisModel}) in the $yz$-plane, where $y$ is the direction of orbital motion (+$y$ for the O star) and $z$ is the orbital axis. The circle of radius $r_c$ is located a distance $x_c$ (blue) from the WN star and is offset a distance $y_c$ along purple line in Figure \ref{fi:ChrisModel} from the line between the stars. Since the system's circular orbit implies constant emission, all phase variation in X-rays is due to changes in the absorption and occultation of the X-ray emitting region. Our model calculates 64 mass column densities $m_c$ (black) along rays from points evenly distributed around the circle to the observer, whose perspective is rotated around the system (in the $xy$-plane) at inclination $i$\,$=$\,78.3$^\circ$ (from the +$z$-axis) to determine the phase dependence of $m_c$.
When emission locations are occulted by the O star (red) or WN star (cyan), we set the column density to infinity, while the $m_c$ of unocculted locations are determined by setting the entire circumstellar environment to be WN wind material (gray).  Therefore our model overestimates the circumstellar absorption in regions that would be occupied by the less dense O-star wind, but also underestimates the absorption column for regions where the line of sight passes through shocked material.

We compute the density $\rho$ of the WN wind from a spherically symmetric, constant-mass-loss, $\beta$\,=\,1 velocity-law wind with parameters from Table \ref{SysParam}; the column depth integral $m_c=\int \rho\,dz'$ (where $z'$ is the direction of the observer) is thus solved in closed form. We then compute optical depths to each point on the circle using a band-appropriate opacity $\kappa(E)$ (i.e.\ energy $E$ is within the waveband), such that the relative X-ray flux is
\begin{equation}\hspace{50pt}\label{eq:ChrisModel}
L_X(\phi,E)=\sum_{i=1}^{64} w(i)\, \text{exp}(-m_{c,i}(\phi)\kappa(E)),
\end{equation}
where $\phi$ is phase and $w(i)$ is the emission weighting of each point on the circle (see below). We choose  $\kappa(E)$ to be consistent with opacities from a WN {\tt windtabs} model \citep{Leutenegger}. Since this model only produces relative X-ray fluxes, we use a linear and constant normalization term to convert equation \ref{eq:ChrisModel} to {\it XMM-Newton} PN cts s$^{-1}$ for model-to-data comparisons (Figure \ref{XrayLCPN}).

The emission from all points around the circle is not necessarily equal; a larger portion of the wind flows perpendicularly to the shock front on the prograde side than on the retrograde side, so the prograde side should emit more strongly in X-rays (see e.g.\ Pittard 2009).  In V444 Cyg the stronger WN wind leads to the shock front wrapping around the O star, so our model sets the emission weighting of the O-star's prograde edge of the circle to $w_{\rm pro}$ times the emission of the retrograde edge $w_{\rm ret}$\,=\,$1$.  For intermediate points $i$, we set the weight $w(i)$ to $w(i)$\,=\,$f(i)(w_{\rm pro}-1)+1$, where $f(i)$\,=\,($y(i)$\,-\,$y_c$+$r_c$)/(2$r_c$)
is the fractional distance along the horizontal line through the emitting circle's center ($y$ direction, purple dashed line in figure \ref{fi:ChrisModel}) from the retrograde edge ($y_c-r_c$) to the prograde edge ($y_c+r_c$).

Our optimal light-curve models are displayed as solid lines in Figure \ref{XrayLCPN} for the soft (top), medium (second panel), and hard (third panel) bands. The parameters for those models are listed in Table \ref{ta:ChrisModel}. Each observed light curve varies more significantly than can be explained by our occultation plus WN-wind--absorption model, so instead of performing formal statistics to determine a best fit to all data points, we obtain an optimal fit for each waveband by attempting to match the main features of each light curve. The soft model light curve reproduces the sought-after minimum around phase 0.05 and peak just after phase 0.6, but our model peak is not as sharp as that of the observed light curve. However, if we fit the more narrow shape from phase 0.5--0.7, we do not reproduce the gradual change in X-rays over the remaining phases. This might indicate that a two-component model would be more appropriate; in this case one circle of emission would reproduce the narrow shape between phases 0.5 and 0.7 after the O star passes in front of the WN star, and the other would produce the more gradual trend seen throughout the orbit. Since there is no {\it a priori} method for combining relative fluxes between two light curves, we do not consider a two-component model here. Another explanation for the observed peak shape near phase 0.6 is that it arises from intrinsic embedded wind shocks (EWS) in the O-star wind. The comoving nature of EWSs leads to them generating soft X-rays. The timing of the peak soft emission is consistent with when the EWS emission suffers the least absorption due to the system's geometry and shock opening angle.

\begin{table}[t]
  \centering
  \caption{Model parameters for the light curve fits in Figure \ref{XrayLCPN}.}
\begin{tabular}{l c c c}
  \hline
  parameter & soft & medium & hard \\
  \hline
  $\kappa$ (cm$^2$/g) & 79 & 1.3 & 0.66 \\
  $s$ ($R_\odot$) & 45 & 8 & 5 \\
  $p$ ($a$) & 0.75 & 0.65 & 0.6 \\
  $o$ ($R_\odot$) & -15 & -1 & 1 \\
  $w_{\rm pro}$  & 1 & 3 & 2 \\
  \hline
\end{tabular}
  \label{ta:ChrisModel}
\end{table}

Our main fitting goals for the medium and hard light curves are to reproduce the timing and depths of the minima at phases $\sim$\,0.55 and 1.0. Our model achieves this by producing strong absorption while the WN star is in front and occultation when the O star is in front. We match the depths of the minima and the timing of the O-star occultation wall, but the minimum from WN-wind absorption occurs slightly before phase 1.0 in this model. We do not reproduce the rapid flux variations outside of the minima (e.g. the apparent drop in flux around phase 0.75 (green) in the medium band and the large variation around phase 0.9 (dark blue) in the hard band), but if these are cycle-to-cycle stochastic variations, then our model cannot be expected to reproduce these features.

Table \ref{ta:ChrisModel} compares our fitting parameters for each waveband and reveals some encouraging results. The opacities in each X-ray waveband agree well with predicted values for a WN star \citep{Leutenegger}. In addition, in our model the size of the emitting circle ($r_c$) decreases as the energy increases, which is expected since harder X-rays should come from the more direct shocks that occur closer to the line of centers. In the medium and hard bands where occultation is necessary, our emitting circles have approximately $r_c$\,$\sim$\,$R_{\rm O}$, while the soft band's size is $r_c$\,$\sim$\,$a$, indicating that this emission arises from a much larger region.

The relative emission weightings of the prograde side of the shock to the retrograde side in our fits are also encouraging. Our medium- and hard-band fits have the prograde side $\sim$2--3 times stronger, which is plausible since the emission comes from small circles in between the stars. These emission weightings are the dominant reasons behind the asymmetries in our model light curves. Our soft-band model light curve shown in Figure \ref{XrayLCPN} does not have any prograde enhancement ($w_{\rm pro}$\,$=$\,1), but the larger circle size makes the model light curve less sensitive to this parameter, so $w_{\rm pro}$\,$\sim$\,2--3 is acceptable. Instead, the soft-band model asymmetry is caused by a shift in the circle location from the line of centers to $y_c$\,=\,$-15R_\odot$, which causes the maximum (minimum) column density occur after superior (inferior) conjunction.

\begin{figure*}
    \centering
    \includegraphics[scale=0.40]{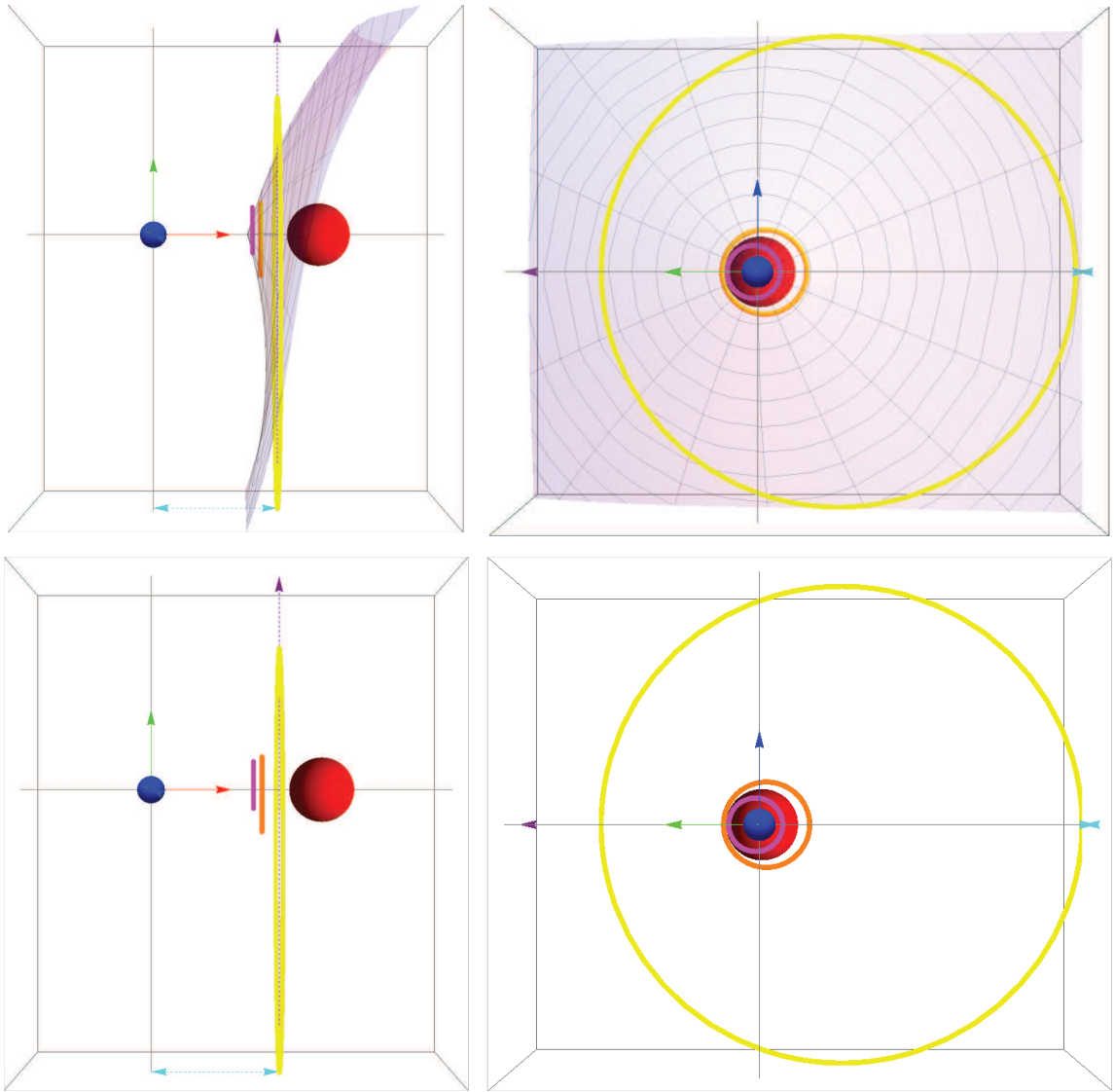}

  \caption{Top view (left) and side view at superior conjunction (right) of the locations and sizes of the optimal circles of emission for the soft (yellow), medium (orange), and hard (magenta) bands.  The red, green, and blue arrows along the axes signify the $x-$, $y-$, and $z-$axes, respectively.  The cyan and purple lines have the same meaning as Figure \ref{fi:ChrisModel}.  The shaded region represents a shock cone with a half-opening angle $\alpha$\,=\,75$^\circ$ and an apex at $p$\,=\,0.57$a$. In the bottom row we remove the 3D structure of shock cone for comparison.}
  \label{fi:ChrisModel3}
\end{figure*}

Figure \ref{fi:ChrisModel3} shows the geometry of our emission circles relative to the stars, with the viewer located above the orbital plane (left) and behind the WN star at superior conjunction (right).  Remarkably, our derived locations of the emission circles appear to trace out a wide-opening-angle shock cone that is wrapped around the O star and skewed by orbital motion.  The shaded regions in Figure \ref{fi:ChrisModel3} show the 3D structure of such a shock cone.  Its half-opening angle is $\alpha$\,=\,75$^\circ$, and its center is located at $\{x,y,z\}$\,=\,$\{0.57a,0,0\}$.  The distortion from orbital motion is computed from an Archimedean spiral; we assume the material travels at the wind speed of the WN star and completes one rotation per orbital period. For better agreement, the soft-band emission circle (yellow) should be rotated several degrees retrograde (clockwise in the left panel). However, our graphic represents only the shock's contact discontinuity; the region of shocked gas should extend out from the contact discontinuity, either due to strong instabilities for a radiative shock, or from an adiabatic shock that naturally has an appreciable thickness \citep{Stevens}. Thus all three emission locations are consistent with a wide-opening-angle shock cone located away from the O star's surface. This is a noteworthy achievement of this model, as it agrees with the prediction of radiative braking and inhibition (Gayley et al.\ 1997, see bottom portion of their Figure 3; Stevens et al.\ 1994). 

\begin{figure*}
    \centering
    \includegraphics[scale=0.70]{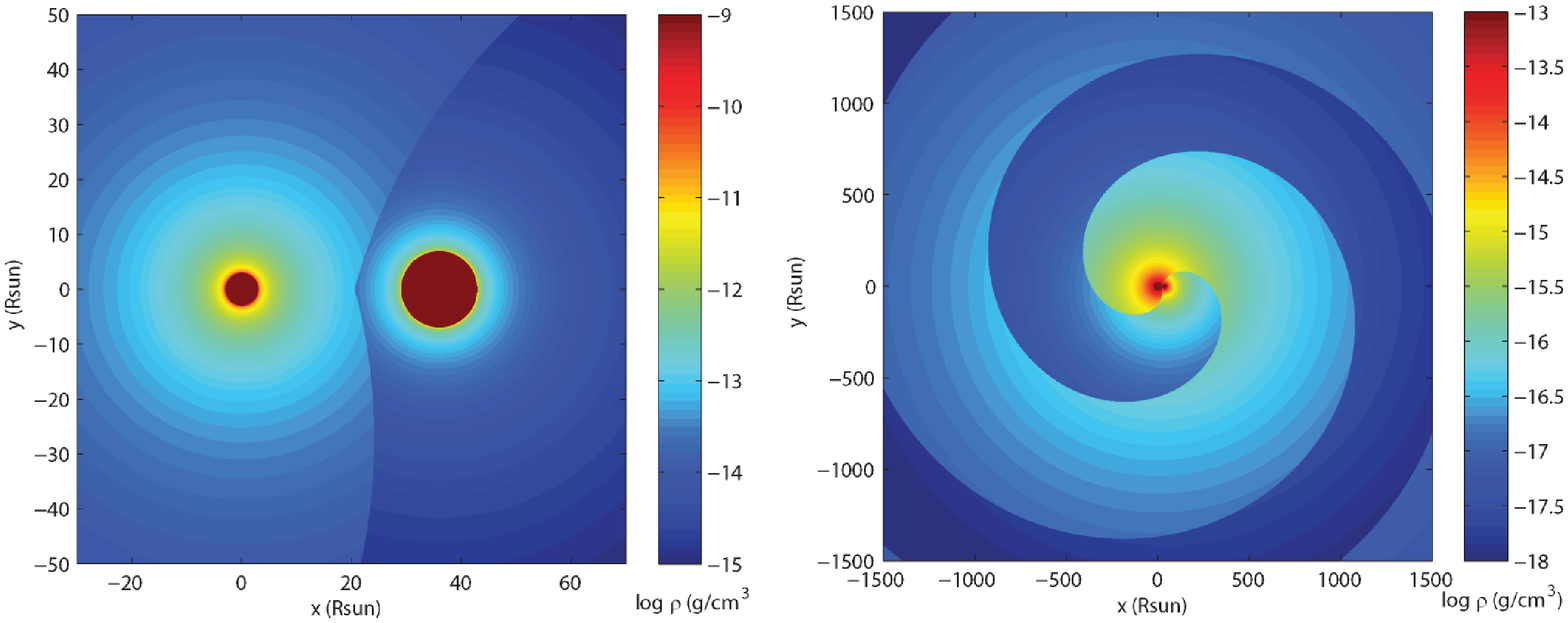}

  \caption{Density in the orbital plane of the two-wind plus shock-cone model on two different spatial scales.  The WN star is on the left and O star is on the right.  The parameters of the shock cone are determined from the occultation plus WN-wind--absorption model.}
  \label{NewModel}
\end{figure*}

Because our occultation plus WN-wind--absorption model includes only WN-wind material, the results it produces for lines of sight looking through the less dense O-star wind may not be valid. To investigate this effect, we also created a two-wind plus shock cone model that incorporates the shock cone found in our previous modeling to separate the circumstellar regions occupied by WN wind and O-star wind. We calculated the densities for each with using the parameters listed in Table 1, assuming for simplicity that the wind material in both regions are spherically symmetric and unperturbed. These assumptions are sure to break down in regions far from the stars since the winds travel at different speeds. However, this is not the case in regions close to the stars, which are the most important for accurately computing column densities. Our resulting two-wind plus shock cone model does not include shocked material; this is not a problem if this material is hot, but the model underestimates column density contributions from farther downstream where the gas has cooled are underestimated. Figure \ref{NewModel} shows the density in the orbital plane of this model.

To compute this model's light curves, we used the same three emitting circles as the previous model, keeping all of their parameters the same (i.e. location, size, prograde weighting, and opacity) except for the constant and linear normalization terms used to scale our model light curve to the data. We also numerically integrated densities from our light of sight to the 64 locations around each circle, because in this case there is no longer an analytic expression for the density along each ray. 

Figure \ref{XrayLCPN} shows our resulting two-wind plus shock-cone model light curves (dashed line). The hard and medium light curves look very similar to those produced by our initial occultation plus WN-wind--absorption model. At phase regions when the lines of sight go through the O-star wind (near phase 0.5), occultation, not absorption, is the dominant effect. The opacities in this region are also low enough that the lower column densities do not cause appreciable variations. In the soft band, by contrast, the model light curve varies more significantly from our previous model due to the model's higher opacity. There are minuscule variations between models when looking through the WN wind, but the drop in column density when looking through the O-star wind increases the X-ray flux for phases around secondary eclipse, thus altering the shape of the soft model light curve. After scaling our models to the X-ray observations (which requires different normalization terms than our occultation plus WN-wind--absorption model) we can compare the peak and trough locations and see that they match the data well.

Given the similarity between the results of the two-wind plus shock cone model with the simpler occultation plus WN-wind--absorption model, the same conclusions still hold. Therefore, our more complex model is also consistent with the emerging picture of V444 Cyg, in which radiative braking or inhibition produces a shock cone with a large opening angle. We discuss the implications of this result in further detail in Section 5.

\subsection{Preliminary polarization modeling}

The phase dependent behavior of the line polarization observed in the V444 Cyg system is quite complex (Section 3.3). In order to properly model the system, several different components need to be taken into account. First, because the system is a binary surrounded by outflowing material, the line polarization should exhibit the approximately sinusoidal variations predicted by \cite{BME}. This is complicated by polarization created by electron scattering in the ionized shock and a distortion of the winds from spherical symmetry \citep{StLouis}. We have conducted preliminary radiative transfer modeling to determine the contribution to the polarization from electron scattering in the shock using a Wilkin-type model \citep{Wilkin}. In this case we assumed line emission arises from the O star and that the WN wind resembles a dense interstellar medium moving at $v^\infty_{WN}=2500$ km s$^{ -1}$. We took the standoff distance between the shock and the O star as the distance implied by the location of the stagnation point from the hard X-ray models (see Section 4.1). This preliminary work suggests that polarization from scattering in the shock alone cannot explain the observed line polarization because it predicts a polarization of no more than 0.1\% whereas our data show polarization on the order of 1\%. The remainder of the polarization is likely due to scattering within the WN wind, which is distorted due to the orbital motion of the system, and the cavity described in Section 3.3. We continue to refine this model to account for eclipse effects, density enhancements within the wind and shock, and emission from sources other than the O star (where our current model assumes it arises), and will report findings in a future contribution. 

\subsection{Comparison with other observational estimates and theoretical predictions}

The expected value of the shock half-angle with no radiative inhibition or braking can be calculated using Equation 3 from \cite{Eichler}
$$\theta \simeq 2.1 \left( 1- \frac{\eta^{\frac{2}{5}}}{4}\right) \eta^{\frac{1}{3}},$$
where $\eta=\frac{\dot{M}_{O6} V^\infty_{O6}}{\dot{M}_{WN} V^\infty_{WN}}\simeq 0.058$ using the system parameters listed in Table \ref{SysParam}. This calculation yields a half-opening angle of approximately $42\degree$ for the shock in V444 Cyg. Similarly, Equation 9 from \cite{Gayley09},$$\eta=\frac{tan(\theta)-\theta}{tan(\theta)-\theta+\pi},$$ leads to an estimated half-opening angle of $44\degree$. Both of these are similar to the $40\degree$ estimate of \cite{Shore}. Equation 28 from \cite{Canto}, $$\theta-tan(\theta)=\frac{\pi}{1-\beta},$$ where $\beta=\frac{\dot{M}_{O6} V_{O6}}{\dot{M}_{WN} V_{WN}}$ yields a half-opening angle of $68\degree$ when we assume both winds have accelerated up to terminal velocity ($\beta=\eta$). Except for the estimate based on the work of \cite{Canto}, these are significantly smaller than the half-opening angle we found for the shock using our simple model (approximately 75$\degree$; Section 4.1). However, our larger value agrees with several other observational estimates. \cite{Flores} found a similar shock half-opening angle of 70$\degree$ using the HeII $\lambda$4686 line, while \cite{Marchenko1994} initially found an opening angle between 82$\degree$ and 108$\degree$ using absorption features in a HeI line and later refined this estimate to 50-60$\degree$ \citep{Marchenko}. Additionally, in their 1997 analysis, Marchenko et al. suggested that one of the bow shock arms crosses our line of sight near phase 0.73. This is very close to the phase at which our 2.0 keV absorption column density increases (Figure \ref{XraySpecParam}) and the polarization behavior of the He II $\lambda$ 4686 and N IV $\lambda$7125 lines changes (Figures \ref{HeFigure} and \ref{NFigure}). Thus our observational results support the large opening angle proposed by these authors.

\cite{Gayley} considered V444 Cyg in their investigation of radiative braking, a process by which the WN star wind may be decelerated through interactions with radiation originating from the O star in the system (see also Stevens et al.\ 1992). In the case of no or low radiative braking or inhibition the WN star wind would collide directly with the O-star surface. Additionally, \cite{Gayley} found that radiative braking can significantly affect both the opening angle of the shock as well as its location. In a case of strong radiative braking, the opening angle of the shock can be large; for example, Figure 3c from \cite{Gayley} shows a case in which the interaction region between the O and WN stars winds is nearly planar (i.e. the shock's half-opening angle is 90$\degree$). Although this study assumed slightly different parameters for the V444 Cyg system than we have assumed here, our observations imply a shock structure offset from the O-star surface (Figure \ref{fi:ChrisModel3}) and a large opening angle, both of which are consistent with the results predicted for strong radiative braking. In fact, the results of more recent modeling work that included relatively weak radiative braking (such that the WN wind collides with the O-star surface) are also partially comparable to our findings \citep{Pittard1999,Pittard2002}. The models have an opening angle of 120$\degree$ (compared to our 150$\degree$) and one of the bow shock arms crosses our line of sight at phase 0.72 (we find phase $\sim$0.75). 

\cite{Shore} found that the projected shock axis should cross our line of sight between phases 0.55 and 0.57, around the same phase as the O star's eclipse of the stagnation point (phase 0.55 from the hard X-ray light curve; Section 3.2). \cite{Pittard1999} and \cite{Pittard2002} found a similar shock axis angle (19$\degree$) for the cone which places the shock axis at about 0.553 in phase. Combining our data and model information together suggests that the shock cone is skewed due to the orbital motion of the stars, which is consistent with these previous findings. We expect that more sophisticated, future modeling will determine the location of the shock cone peak and level of distortion in a more precise manner.


\section{Summary}

We have presented a new set of X-ray observations of V444 Cyg which we combined with archival observations to create the most complete set of X-ray light curves of the system to date. We also presented and analyzed supporting optical spectropolarimetric data, creating polariztion phase curves of three strong emission lines.

The soft X-rays show low absorption, and are therefore arise relatively far from the stars. The soft X-ray count is highest when the less dense wind of the O star lies directly in our line of sight (around phase 0.5). This results from a combination of two effects: lower absorption in the O-star wind than in the WN wind and additional soft X-ray emission arising within the O-star wind itself. 

The hard X-rays are arise in the wind-wind collision region between the two stars. Therefore, they suffer a higher absorption when the WN star is in front (around phase 0.0), and are lower one when the O star is in front. The relative duration of these eclipses implies a large opening angle of the shock cone consistent with radiative braking and inhibition predictions for V444 Cyg. Additionally, we detect eclipses of the wind-wind collision region by the stars. However, the fact that the eclipses are not total suggests that the X-ray emitting part of the wind-wind collision region is larger than either star. 

We have also directly detected the effects of Coriolis distortion within a wind-wind collision shock for the first time in X-rays. The effect manifests itself through asymmetries in the soft light curve and the timing of eclipses of the hard X-rays, which are shifted with respect to the optical eclipses.

Our polarization results suggest that the shock and O-star wind create a cavity of missing material in the HeII and NIV shells around the WN star. We found that the line polarization behavior of the system is not similar to that of the continuum. This suggests that the scattering regions for line and continuum photons within the system are not the same. Additionally, our analysis of their variations suggests a large opening angle for the shock.

To interpret our X-ray data, we created a simple model of the system, using only occultation and absorption effects. This allowed us to reproduce the major features of the X-ray light curves and showed that the shock cone must have a large opening angle consistent with radiative braking and inhibition. 

Taken together, our X-ray and spectropolarimetric data sets reveal that both Coriolis distortion and radiative braking or inhibition must be present within the system to explain the location of the shock and its large opening angle.

\begin{acknowledgements}

This work is based on observations obtained with \textit{XMM-Newton}, an ESA science mission with instruments and contributions directly funded by ESA Member States and NASA. We are very grateful to Brian Babler, Karen Bjorkman, Marilyn Meade, and Ken Nordsieck for their help with the HPOL data and to all of the members of the PBO science team for obtaining the HPOL data used in this paper. Yoshitomo Maeda was helpful in obtaining our new \textit{XMM-Newton} observations and Richard Ignace provided valuable comments on our manuscript. We thank Ken Gayley for his helpful referee comments. We acknowledge support through NASA ADP award NNH12ZDA001N. JRL would like to acknowledge support from the NASA Harriett G. Jenkins Pre-doctoral Fellowship Program and Sigma Xi's Grants-in-Aid of Research Program. Additionally, JRL is grateful for the hospitality of the D\'{e}partement AGO at the Universit\'{e} de Li\`{e}ge for hosting her during the time this work was completed. YN acknowledges support from the Fonds National de la Recherche Scientifique (Belgium), the PRODEX \textit{XMM-Newton}, and \textit{Integral} contracts. JLH is grateful for support from NSF grant AST-0807477.
      
\end{acknowledgements}


\end{document}